\documentclass[]{article}
\usepackage{multicol}
\usepackage{apj}
\newcommand {\ks} {${\rm km~s}^{-1} \;$}

\newcommand {\m} {\mbox{$\mu$m }}
\newcommand {\hnot} {\mbox{${\rm H}_0$}}

\begin{document}

\vspace{15mm}
\begin{center}
\uppercase{
The Near- and  Mid-Infrared Continuum Emission of Seyfert Nuclei: 
Constraints on the Models of Obscuring Tori}\\
\vspace*{1.5ex}
{\sc 
Dario Fadda$^{1,2,4}$,  Giuliano  Giuricin$^{1,2}$, Gian Luigi Granato$^{3}$ and Donatella Vecchies$^{1,2}$ 
}\\
%\vspace*{-1pt}
\vspace*{1.ex}
{\small 
$^1$Dipartimento di Astronomia, Universit\`{a} degli Studi di Trieste, \\
$^2$SISSA, via Beirut 4, 34013 -- Trieste, Italy\\ 
$^3$ Osservatorio Astronomico di Padova, Vicolo dell'Osservatorio 5, 35122 -- Padova, Italy\\
$^4$ Osservatorio Astronomico di Trieste, Via Tiepolo 11, 34131 --Trieste, Italy\\
E-mail:
fadda@sissa.it; giuricin@sissa.it; granato@pdmida.pd.astro.it; dona@sissa.it}
\end{center}
\vspace*{-6pt}

\begin{abstract}
For an extended sample of Seyfert galaxies we compile from the literature
the infrared fluxes in the four IRAS bands, the ground--based small--beam 
($\sim$5-10") fluxes in the standard Q, N, M , L (or L') bands, and the
nuclear (non-stellar) estimated fluxes in the JHK bands. We estimate
nuclear fluxes in the L band by applying a correction for stellar light. 

From the statistical study of the infrared colors and luminosities, we
derive the typical SEDs of Seyfert 1 and 2 nuclei and the typical
differences in luminosities between the two types of objects in the mid-
and near-infrared spectral ranges. The analyses of colors and luminosities
agree with the fact that in general Seyfert 2 nuclei become increasingly
fainter with respect to Seyfert 1 as we go toward shorter infrared
wavelengths. This behavior is consistent with growing anisotropy of a
dusty torus emission towards shorter wavelengths, but the degree of
anisotropy is low (the radiation appears to be substantially isotropic at
$\lambda\gtrsim$25 \m). For Seyfert 2 galaxies having Compton--thin
obscuring structures at hard X-ray energies, we find correlations between
the absorbing hydrogen columns and some infrared colors and luminosities. 
  
The observational data appear to severely challenge many models of dusty
tori, which hardly account for the shapes of the SEDs and the degree of
anisotropy observed in Seyfert galaxies. In particular, at variance with
some earlier claims, very thick and compact tori are basically
inconsistent with these observational constraints. The most successful
models, though having problems in accounting for several details, can
fit the major infrared observational data of both Seyfert 1 and Seyfert 2
nuclei with tori which extend up to several hundreds pc and have fairly
low optical thickness. \\
\vspace*{-6pt}
\noindent
{\em Subject headings: } 
galaxies: active -- galaxies:nuclei --  galaxies: Seyfert -- infrared: galaxies
\end{abstract}  
\begin{multicols}{2}

\section{Introduction}

Large amounts of observational data have been accumulated during the last
decade supporting unified models of active galactic nuclei (AGNs),
particularly of Seyfert galaxies. The canonical unified model of Seyfert
galaxies claims that Seyfert 1 galaxies (S1s) contain the same central
engines as Seyfert 2 galaxies (S2s), but have the broad-line region and
the strong optical/UV/X--ray continuum of the central source concealed
from our view by a disk or torus of dusty molecular clouds (see, e.g., the
review by Antonucci, 1993). This obscuring structure would absorb a
significant fraction of the optical/UV/X-ray luminosity and should
therefore radiate this energy strongly in the infrared (IR). This would
explain why Seyfert galaxies are strong IR emitters. They appear to be
relatively brighter than non--Seyfert ones at near--infrared (NIR) and
mid--infrared (MIR) wavelengths on the basis of the global spectral energy
distributions (SEDs) and relevant central colors (e.g., Spinoglio et al.,
1995; Glass \& Moorwood, 1985; Giuricin et al., 1994). 
  
Dust reprocessing of the primary emission is at present regarded as the
most likely source of the bulk of the strong NIR and MIR emission of
radio--quiet quasars and Seyfert galaxies. Observational evidence
favoring models in which the IR emission predominantly or entirely comes
from heated dust is summarized by Barvainis (1992). 

Geometric, energetic, and spectropolarimetric arguments indicate that the
optical/UV continuum escapes anisotropically from the nuclei of at least
some Seyfert objects. The presence of extended ionization cones which are
seen in images of emission lines around several S2 galaxies (see, e.g.,
the review by Wilson, 1996) suggests that the ionizing radiation emitted
from the nucleus is collimated. The observed ionizing photon deficit in
S2s (Binette, Fosbury \& Parker, 1993) indicates that the central ionized
gas responds to a higher photon flux than that seen along our line of
sight.  Broad hydrogen emission lines, presumably transmitted through the
torus, have been detected in the IR spectra of some S2 galaxies. (Blanco,
Ward \& Wright, 1990; Goodrich, Veilleux \& Hill, 1994; Ruiz, Rieke \&
Schmidt, 1994). These kinds of arguments suggest the presence of an
obscuring dusty torus in S2 objects (see,e.g., the review by Wilson (1992)
for a discussion of these topics).  The X--ray spectra of S2s exhibit
absorptions larger than those of S1s, providing strong support for the
presence of an obscuring structure (e. g., Nandra \& Pounds, 1994;  Smith
\& Done, 1996). 

The first models of thermal dust emission for the origin of the IR
emission of Seyfert galaxies and radio-quiet quasars simply involve
optically thin dust configurations (e.g., Barvainis, 1987, 1990, Loska,
Szczerba \& Czerny, 1993), which imply too small an optical depth at
optical/UV wavelengths ($\tau \lesssim 3$), or configurations in idealized
slab geometry (Laor \& Draine, 1993). Phinney (1989; see also Sanders et
al., 1989) modeled the IR continua of AGNs as thermal emission from a
dusty warped disk which extends for a few kpc in the host galaxy; but its
large radial extent would imply an unreasonably large mass, if the warped
structure were not transparent enough at IR wavelengths. 

In recent years radiative transfer calculations specific to the IR
emission of thick, dusty axisymmetric torus-like structures have been
carried out by Pier \& Krolik (1992, 1993, hereafter PK), Granato \&
Danese (1994; hereafter GD), Efstathiou \& Rowan--Robinson (1995;
hereafter ER).  All these models concur that face-on systems (i.e., S1s
within the unified scheme of Seyfert galaxies) should be, on average,
stronger NIR and MIR emitters than edge-on systems (i.e., S2s) for a wide
range of torus geometries and optical depths, whilst at long enough IR
wavelengths the two types of objects should appear to be similar. 
However, the predicted IR emission and degree of IR anisotropy differ in
many details. 

As is expected in many models of optically thick tori, the observations
indicate that Seyfert galaxies display a broad IR bump which peaks in the
MIR. S2s exhibit steeper IR continua between NIR and far--infrared (FIR)
wavelengths than S1s do (Edelson, Malkan \& Rieke, 1987), although an
accurate determination of the SEDs of the two types of objects is hampered
at NIR wavelengths by appreciable starlight contributions, which are
typically greater for S2s (Alonso-Herrero, Ward \& Kotilainen). Moreover,
some authors have claimed that S2s are characterized by lower mean N-band
($\lambda \sim$10 \m) luminosities than those of S1s (Heckman 1995; 
Maiolino et al., 1995). The last two facts can be interpreted as
anisotropy resulting from the presence of an obscuring torus. 
 
Furthermore, the observations show that the MIR spectra of S2s typically
display strong silicate absorption bands (at $\lambda \sim$10 \m),
whereas those of S1 are mostly featureless (Roche et al., 1991). The 
elusiveness of emission features in the MIR spectra of Seyfert galaxies 
represents a serious challenge to many models of tori.
 
Comparisons between IR observations and theoretical predictions have
been presented in several above-mentioned papers for fairly small samples
of Seyfert galaxies (see also Granato, Danese \& Franceschini, 1997). But
the IR SEDs predicted by torus models and, in particular, the level of IR
anisotropy deserve to be better tested through an extensive statistical
analysis of the IR properties of Seyfert galaxies. Their NIR and 
especially MIR continua, where the peak of the torus emission is expected 
to occur, are key spectral regions in which to probe the structure of the
obscuring--reprocessing torus. 

In this paper, in order to test models of IR torus emission, we
analyze the MIR and NIR photometric data of a wide sample of Seyfert
galaxies. Specifically, we compile from the literature IRAS fluxes (at
$\lambda\sim$12, $\sim$25, $\sim$60, $\sim$100 \m), ground-based
small--aperture ($\sim$5--10") photometric measurements in the standard Q
($\lambda \sim$20 \m), N ($\lambda \sim$10 \m), M ($\lambda \sim$4.8 \m),
L ($\lambda \sim$3.5 \m) or L' ($\lambda \sim$3.8 \m) bands, and available
non--stellar (nuclear) flux estimates in the J ($\lambda \sim$1.25 \m), H
($\lambda \sim$1.6 \m), K ($\lambda \sim$2.2 \m) bands. 

In \S 2 we describe the data sample compiled from the literature. In \S 3
we analyze the MIR, NIR colors and luminosities of our Seyfert galaxies
with the main purpose of establishing the relevant SEDs of S1 and S2
nuclei and the differences in their typical relevant luminosities. We also
explore relations between these properties and some hard X-ray properties.
We use survival analysis techniques (see, e. g., the review by Feigelson,
1992) in order to exploit the information contained in censored data
(i.e., upper limits on fluxes). In practice we use the software package
ASURV (Rev 1.2, La Valle, Isobe \& Feigelson, 1992), in which the methods
presented in Feigelson \& Nelson (1985) and in Isobe, Feigelson \& Nelson
(1986) are implemented. Ensuing constraints on the models of IR torus
emission are discussed in \S 4. Lastly, \S 5 summarizes our results and 
contains our conclusions.

\section{The Data Sample}

We have considered a wide sample of Seyfert galaxies ($\sim$700 objects) 
having $z<0.1$. Our sample essentially contains all known Seyfert galaxies  
having $z<0.05$ and many more distant objects. The objects 
have been taken from the lists of Dahari \& De Robertis (1988), Whittle 
(1992), de Grijp et al. (1992), Nelson \& Whittle (1995), and the 
catalogue of V\'eron--Cetty \& V\'eron (1996). Like several wide samples 
of Seyfert galaxies considered in the literature, our sample is likely to be 
biased towards objects which are bright at optical and infrared wavelengths.

The Seyfert type classification is primarily based on the spectroscopic
study of the CfA Seyfert sample by Osterbrok \& Martel (1993) and on the
first four above-mentioned papers. The Seyfert types are grouped into the
three types S1 (types 1 and 1.2), S1.5 (types 1.5) and S2 (types 1.8, 1.9,
2). The distances D of the nearby galaxies which are included in Tully's
(1988) NBG catalog are taken directly from the values tabulated therein.
The distances of non-cluster NBG galaxies have been essentially estimated
on the basis of velocities, an assumed \hnot= 75 $km\ s^{-1} Mpc^{-1}$ and
a Virgocentric retardation model in which the author assumed that the
Milky Way is retarded by 300 $km\ s^{-1}$ from the universal Hubble flow by
the mass of the Virgo cluster. Non-NBG galaxies are given redshift
distances D with the same value of \hnot. 

We have collected the IRAS fluxes (at $\lambda\sim$12, $\sim$25, $\sim$60,
$\sim$100 \m) of our Seyfert objects preferentially from specific
reference sources which report co-added survey data (e.g., Rush, Malkan \&
Spinoglio (1993), Sanders et al. (1995), de Grijp et al. (1992), Edelson
et al. (1987), Sanders et al. (1989), Mazzarella, Bothun \& Boroson
(1991), Knapp, Bies \& van Gorkom (1990), Knapp et al. (1989), Soifer et
al. (1989), Bothun et al. (1989), Hill, Becklin \& Wynn--Williams (1988),
Rice et al (1988), Helou et al.  (1988), Salzer \& MacAlpine (1988),
Golombek, Miley \& Neugebauer (1988)), and from the IRAS Faint Source
Catalog (see Moshir et al., 1992), in order of preference.  We have found
IRAS data (detections or 3 $\sigma$ upper limits) at $\lambda \sim$12, 
$\sim$25, $\sim$60, $\sim$100 \m for 535, 538, 540, 539 objects, 
respectively. 
 
We have considered the recent estimates of the non-stellar (nuclear)
near-infrared (NIR) J, H, K fluxes. We have taken relevant NIR data
primarily from the observational studies by Kotilainen et al.
(1992a,b), who observed a hard X-ray--selected sample of Seyfert galaxies
(nearly all S1), Danese et al. (1992) and Zitelli et al. (1993), who
observed Markarian S1 and S1.5 objects, Kotilainen \& Prieto (1995), who
studied NGC 5252, and Alonso-Herrero, Ward \& Kotilainen (1996), who
investigated S2 objects. From NIR imaging observations (often complemented
by NIR aperture photometry) these authors obtained accurate estimates of
the contribution to the total NIR emission arising from the nuclear source
alone by decomposing the surface brightness profiles into a
combination of a nuclear source, a bulge component and a disc component.
For NGC 5548 and Mk 509 we have adopted the averages of the NIR nuclear
fluxes derived by Kotilainen et al. (1992a) and Danese et al.  (1992). The
resulting data sample contains nuclear J, H, K fluxes for 73 (41 S1, 12
S1.5, 21 S2), 73 (41 S1, 12 S1.5, 21 S2), and 79 (46 S1, 12 S1.5, 21 S2)
objects respectively.  Moreover, we have taken some L-band nuclear
magnitudes estimated by Alonso-Herrero et al. (1996) through model
parameters obtained in the K band, L aperture photometry and standard K-L
colors of normal galaxies. 

We have gathered other (less accurate) K-band nuclear data from McLeod \&
Rieke (1995) and from Heisler, De Robertis \& Nadeau (1996) for 24 and 15
objects respectively. In their own K-band imaging observations of CfA
Seyfert objects the former authors have approximated the nuclear source as
a stellar point spread function scaled to the central intensity of the
light profile; this approach tends to overestimate the nuclear source
contribution. Heisler et al. (1996) have reported NIR images of a sample
of IRAS galaxies whose SED peaks in the IRAS $\lambda \sim$60 \m band and
have derived the nuclear K-band magnitudes of their Seyfert objects from the
observed $H\alpha$ luminosities simply by assuming Case B recombination
and a power--law SED with a spectral index $n=-1$ ($F_\nu \propto \nu^n$) in
the optical and NIR bands. 

In addition, a few JHKL nuclear data have been taken from Ward et al.
(1987) for NGC 4051 and from McAlary \& Rieke (1988) for Mk 6, Mk 1040,
NGC 2992, NGC 6814, I 4329A (for which mean values have been adopted).
Ward et al. (1987) have fitted the light profile obtained from R--band CCD
images with three components (point source, disk, and bulge) and have then
extrapolated the derived nuclear contribution into the infrared assuming
normal stellar colors.  McAlary \& Rieke (1988) employed a Hubble--law
growth curve, modified by a color--morphological type--aperture relation,
to deconvolve multi--aperture optical and NIR photometry of Seyfert
objects into stellar and non--stellar nuclear emissions. 

Moreover, we have gathered together from the literature all published
ground-based, small-aperture ($\sim$5"--10") photometric measures in the
Q, N, M, L or L' standard broad bands. L' band measures are considered
only for objects which lack small-beam L photometry. The consultation of
the "Catalog of Infrared Observations" by Gezari et al. (1993) has been of
valuable aid in compiling infrared photometric data from the literature.
In the case of multiple entries for an object, we have adopted the mean
values of the fluxes published in a reference source. Non--detections are
reported as 3 $\sigma$ upper limits on fluxes; in case of no detection we
have chosen the most stringent upper limits on fluxes. 

In general we have preferred the smallest beam size measures, in order to
minimize the inclusion of starlight in MIR fluxes.  In any case, we have
preferentially selected the sets of multi-wavelength observations
published by the same authors (e.g., the sets of measures by Rieke (1978),
McAlary et al. (1983), Lawrence et al. (1985), Rudy \& Rodriguez--Espinosa
(1985), Edelson et al.(1987), Ward et al. (1987), Neugebauer et al.
(1987), Hill et al. (1988), in order to minimize the effects of data
inhomogeneity and infrared variability (if any) in the definition of the
spectral energy distributions (SEDs) of our objects. Long time-scale 
variations of JHKL band fluxes have been frequently observed in Seyfert  
galaxies (e.g., Kotilainen et al., 1992b; Glass, 1992), whereas no   
sure evidence of variability at longer infrared wavelengths has been 
noted in Seyfert galaxies and radio--quiet quasars (e.g., Edelson \&
Malkan, 1987; Neugebauer et al., 1989). Thus, we have paid particular 
attention to selecting L (or L') fluxes and JHK nuclear fluxes 
measured at the same epoch (if available). For this reason we have 
selected the JHKL nuclear fluxes of NGC 2992 and I 4329 A as given 
by McAlary \& Rieke (1988) (instead of Kotilainen et al.'s (1982a) values),
because they are based on small-beam (4.6") JHKL photometric observations 
which are simultaneous (McAlary et al., 1983). 
 
Most of the QNML(L') photometric data have been taken from the papers
mentioned above and from Stein \& Weedmann (1976), McAlary, McLaren \&
Crabtree (1979), Rudy, Levan \& Rodriguez--Espinosa (1982), Glass \&
Mooorwood (1985), Boisson \& Durret (1986), Carico et al. (1988), Sanders
et al. (1989), Kotilainen et al. (1992), Zitelli et al. (1993),
Wynn-Williams \& Becklin (1993), Vader et al. (1993), Elvis et al. (1994),
Maiolino et al. (1995), and Alonso-Herrero et al.  (1996). We have found 
Q, N, M, L (or L') data (detections or upper limits) for 45 (20 S1, 5 
S1.5, 20 S2), 173 (65 S1, 17 S1.5, 91 S2), 36 (14 S1, 7 S1.5, 15 S2),   
and 202 (79 S1, 21 S1.5, 102 S2) objects, respectively. 

If magnitudes are given in the literature, they have been generally
converted to fluxes by means of Wilson et al.'s (1972) zero--magnitude
calibration.  In view of the fairly large photometric errors (i.e.,
$\sim$10\%, $\sim$10\%, $\sim$20\%, $\sim$20\%, $\sim$25\% for the L, L',
M, N, Q fluxes), no corrections have been applied for interstellar
extinction and redshift. 

Finally, we have considered the measures of [OIII]$\lambda$5007A narrow
emission line flux available in the literature for 364 objects (105 S1, 35
S1.5, 224 S2), as compiled mostly by Nelson \& Whittle (1995), Whittle
(1992), Dahari \& De Robertis (1988), de Grijp et al. (1992), Mulchaey,
Wilson \& Tsvetanov (1996a). 

The starlight contribution dramatically decreases as we go from the 
JHK bands to longer wavelengths (see, e.g., the JHKL nuclear fluxes 
estimated by Ward et al. (1987) and McAlary \& Rieke (1988)). Also 
the image sizes of luminous infrared galaxies are smaller at 
$\lambda \sim$3.4 \m that at shorter IR wavelengths (e.g., Zhou, 
Wynn--Williams \& Becklin, 1993). The small--beam fluxes in the QNM 
bands of our Seyfert objects can be taken essentially as nuclear fluxes.

On the  other  hand,  the stellar  contribution is  still  appreciable
within  small--beam L   fluxes; within  3 arcsec it  ranges   between
fractions of $\sim$10\% to $\sim$50\% (with  a mean of $\sim$30\%) for
S2s, which, however, show a greater starlight contribution in the JHKL
bands than S1s (Alonso-Herrero  et   al., 1996). Therefore, we    have
corrected  the published L (or  L') fluxes by subtracting an estimated
value for  starlight. This was done  by assuming that the J-band flux,
which has been  taken in the same  aperture and preferentially by  the
same  authors as the L (or  L')  measurements,  consists entirely of
starlight which has  J$-$L=+1.3; this is  the normal color for an old
stellar population coming from observations of elliptical galaxies and
bulge--dominated regions of  spiral galaxies  (e.g., Willner  et  al.,
1984).  Any superposing thermal  emission from central hot dust  would
enhance the L-band  flux much more than  the  J-band flux so that  the
resulting overcorrection in L would be negligible.

Since a few galaxies lack J-band photometry, the resulting sample of
galaxies with L-band nuclear fluxes or relevant upper limits comprises 200
objects (78 S1, 21 S2, 101 S2). We have checked that our rough method
would give L nuclear magnitudes similar to those less crudely
derived by Alonso-Herrero et al. (1996) for 11 galaxies (in this
case we directly took their values).  In a few cases our L-band correction
is slightly greater than the measured L-band fluxes. For these galaxies,
which have probably very little or no L-band flux associated with an
active nucleus, we have adopted an upper limit of $F_{L_n}<1$ mJy. 

The large--beam IRAS MIR fluxes at $\sim$12 and $\sim$25 \m certainly
include emission coming from the host galaxies.  However, their good
correlations with the K nuclear emission and the N emission suggest that
the bulk of IRAS MIR emission of Seyfert galaxies comes from the nuclear
region, at variance with the IRAS FIR emission (at $\lambda \sim$60 and
$\sim$100 \m) which mostly originates from the host galaxies (see, e.g.,
Danese et al., 1992;  Giuricin, Mardirossian \& Mezzetti, 1995).
This is also directly revealed by the high-resolution IR images of the central 
regions of nearby Seyfert galaxies, since these observations clarify
 the relative role of AGN and star formation in
producing the IR emission (see e.g. the review 
by Telesco (1988) and the MIR maps of NGC 1068 (Telesco \& Decher,
 1988), NGC 1808, NGC 1365, and NGC 5506 (Telesco, Dressel \& Wolstencroft,
 1993), NGC 7469 (Miles, Houck \& Hayward, 1994)).
Therefore, in this paper we mainly deal with the IRAS MIR fluxes. 
We use the IRAS FIR fluxes only for some comparisons.

Throughout this paper the fluxes measured in the IRAS $\lambda \sim$25 and
$\sim$12 \m , Q, N, M bands and the non--stellar (nuclear) estimated
fluxes in the L, K, H, J bands are respectively denoted by $F_{25}$,
$F_{12}$, $F_Q$, $F_N$, $F_{M}$, $F_{Ln}$, $F_{Kn}$, $F_{Hn}$, $F_{Jn}$.
The corresponding monochromatic luminosities $L_\nu$ (expressed in W/Hz)
and the luminosity $L_{[OIII]}$ (in erg/s) of the [OIII]$\lambda$5007 
\\A emission line follow from the adopted distances (with \hnot=75 \ks 
Mpc$^{-1}$). 
We have always used L'-band measures in place of L-band measures, 
if the latter are not available. We have verified that S1 objects 
alone give results (color and luminosity distributions) similar 
to those of S1+S1.5 objects. Therefore, in the following sections the 
latter objects will be simply denoted as S1s. 

\section{Analysis and Results}

\subsection{The IR colors}

In order to establish the behavior of the IR continuum at MIR and NIR
wavelengths, we have evaluated the following colors (decimal logarithm of
flux ratios) $\log (F_{25}/F_{12})$, $\log (F_Q/F_N)$, $\log (F_N/F_M)$,
$\log (F_M/F_{Ln})$, $\log (F_N/F_{Ln})$, $\log (F_N/F_{Kn})$, $\log
(F_{Ln}/F_{Kn})$, $\log (F_{Kn}/F_{Hn})$, $\log (F_{Hn}/F_{Jn})$ separately
for S1+S1.5 and S2 galaxies.  We have evaluated the nine above--mentioned
colors (or relevant lower or upper limits) for all objects detected in
the two relevant bands and also for the objects having upper limits on
$F_{12}$, $F_Q$, $F_M$, $F_M$, $F_N$, $F_N$, for the first six 
colors, respectively. The very few objects having upper limits on the
other term of a given ratio or those which have upper limits on both terms
are excluded. There are only detections for the case of the last three 
colors. 

We have employed the Kaplan--Meier product limit estimator (which
is part of the ASURV package) in order to calculate the median, mean and
distribution functions of the colors of the two groups of Seyfert
galaxies. Redistributing upper limits (if any) uniformly along all the
intervals of lower detected values, the Kaplan--Meier estimator provides
an efficient, non--parametric reconstruction of information lost by
censoring in the case of randomly censored data sets. 
We compare the distribution functions of colors for the two classes of
Seyfert galaxies by testing the "null" hypothesis that the two
independent random samples are randomly drawn from the same parent
population.  In the case of uncensored data, this is accomplished by
using the Kolmogorov-Smirnov test (see e.g. Hoel, 1971).
In the general case of censoring we use two versions of Gehan's
test (one for permutation variance and the other with hypergeometric
variance), the logrank, the Peto--Peto, and Peto--Prentice
tests.\\
%%FIGURE 1%%%
\end{multicols}
\includegraphics{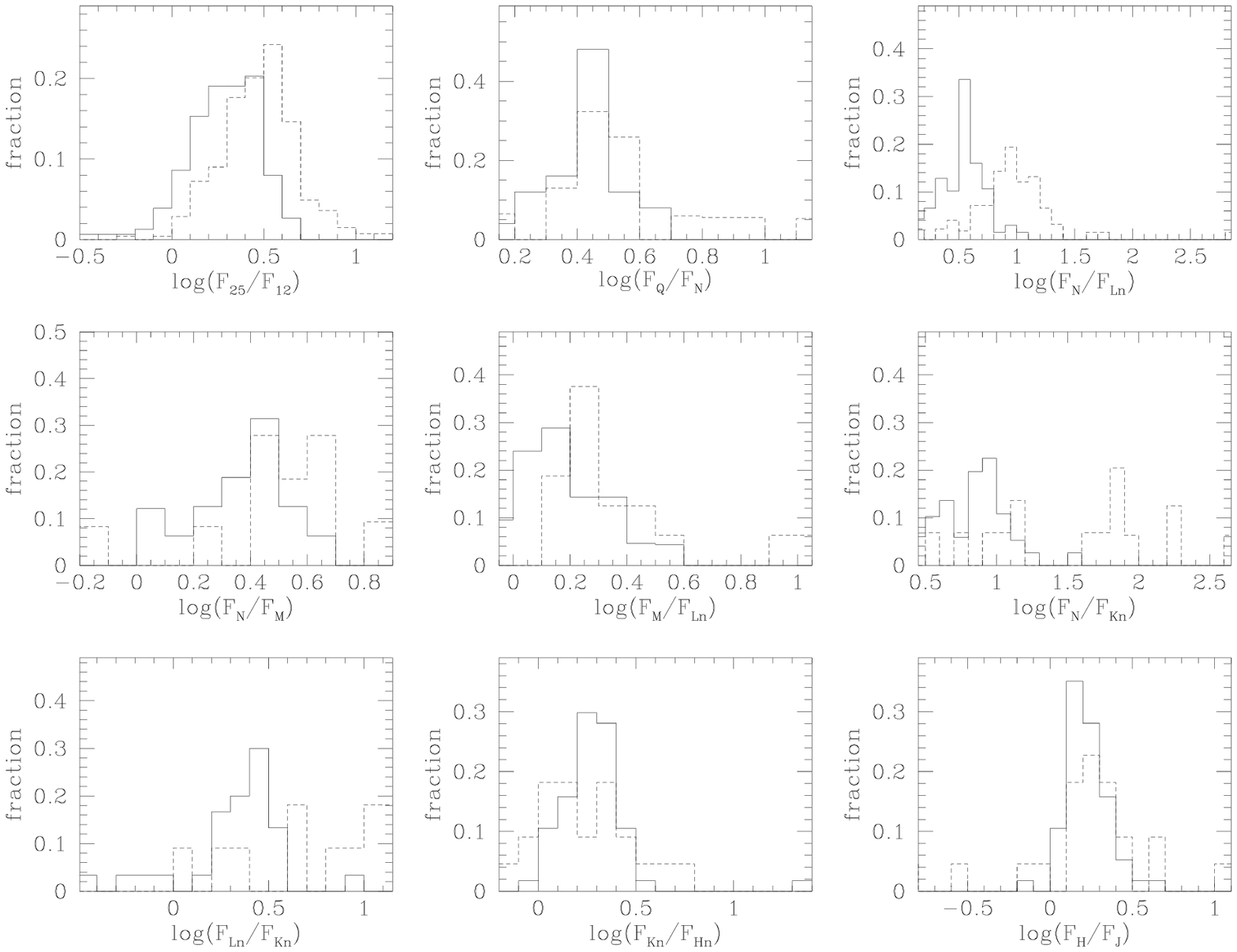}
$\ \ \ \ \ \ $\\
\vspace{13.4cm}
$\ \ \ $\\
\vspace{1mm}
{\small\parindent=3.5mm
{\sc Fig.}~1.---
The histograms show the color distributions for S1
(solid lines) and S2 nuclei (dashed lines).
}
\vspace{5mm}
\begin{multicols}{2}

 These two--sample tests give the probability $p$ that the two
data samples would be drawn from one parent population. The
statistical significance of the difference between the color
distributions of the two classes of galaxies is at the $100 \ (1-p)\%$
level.  The two--sample tests differ in how the censored points are
weighted and consequently have different sensitivities and
efficiencies with different distributions and censoring patterns.\\

\subsubsection{The color distributions}

Table  1  gives the numerical   outcomes,  namely the  total number of
objects N,  the number $N_{ul}$ of  relevant upper  limits, the median
and  mean  values (with     associated 1 $\sigma$   errors),   and the
corresponding median spectral index  $n$ ($F_{\nu}  \propto \nu^{-n}$)
for the distributions of the above--mentioned nine colors.  
Fig. 1 shows the Kaplan--Meier distributions of the colors for S1 and
S2 objects.

In general  S2s have  redder $\log(F_{25}/F_{12})$,  $\log (F_Q/F_N)$,
$\log  (F_N/F_M)$,  $\log  (F_M/F_{Ln})$, $\log  (F_N/F_{Ln})$,  $\log
(F_N/F_{Kn})$, $\log (F_{Ln}/F_{Kn})$ colors (i.e., steeper SEDs) than
S1s.    
Table 2 contains the mean and median values of the probability $p$ that
the colors of S1 and S2 objects would be drawn from the same parent
population. One can see that the statistical significance of  the color
differences between the   two groups of galaxies  is  high  in several
cases   and marginal   in the   cases  of  $\log (F_Q/F_N)$ and  $\log
(F_N/F_M)$.  On the  other hand, the  two groups of Seyfert objects do
not differ  appreciably    in the $\log (F_{Kn}/F_{Hn})$    and  $\log
(F_{Hn}/F_{Jn})$  distributions.

We have attempted to investigate the possible effects of circumnuclear
star formation on IR colors.  Although it is difficult to come up with
consistent quantitative indicators of this phenomenon, it does not seem
to occur   very frequently  in  nearby  Seyfert  objects (e.g., Pogge,
1989).\\
%%TAB1
\vspace{4mm}
\hspace{-4mm}
\begin{minipage}{9cm}
\renewcommand{\arraystretch}{1.2}
\renewcommand{\tabcolsep}{1.2mm}
\begin{center}
\vspace{-3mm}
TABLE 1\\
\vspace{2mm}
{\sc The IR Colors\\}
\footnotesize
\vspace{2mm}
\begin{tabular}{lcrrrrr}
\hline \hline

\multicolumn{1}{c}{Color}       & \multicolumn{1}{c}{Sample}  & 
\multicolumn{1}{c}{$N$}         & \multicolumn{1}{c}{$N_{ul}$}& 
\multicolumn{1}{c}{Median}      & \multicolumn{1}{c}{Mean}    &
\multicolumn{1}{c}{$n$}  
\\
\multicolumn{1}{c}{(1)}       & \multicolumn{1}{c}{(2)}    & 
\multicolumn{1}{c}{(3)}       & \multicolumn{1}{c}{(4)}    & 
\multicolumn{1}{c}{(5)}       & \multicolumn{1}{c}{(6)}    &
\multicolumn{1}{c}{(7)} \\
hline 

$\log (F_{25}/F_{12})$& S1& 157& 26& 0.31$^{+0.0}_{-0.04}$& 0.28$\pm$ 0.02&1.0 \\
                    & S2     & 276& 77& 0.49$^{+0.01}_{-0.01}$&0.47$\pm$ 0.02&1.5\\
$\log (F_Q/F_N)$& S1& 25& 0& 0.42$^{+0.04}_{-0.01}$& 0.42$\pm$ 0.02&1.4\\
              & S2     & 19& 3& 0.47$^{+0.11}_{-0.03}$& 0.55$\pm$ 0.06&1.6\\
$\log (F_N/F_{Ln})$& S1& 67& 8& 0.55$^{+0.01}_{-0.03}$& 0.54$\pm$ 0.02&1.2\\
              & S2     & 66& 9& 0.93$^{+0.03}_{-0.03}$& 0.94$\pm$ 0.05&2.0\\
$\log (F_N/F_{M})$& S1& 18& 2& 0.45$^{+0.03}_{-0.09}$& 0.37$\pm$ 0.04&1.4\\
              & S2     & 12& 1& 0.57$^{+0.06}_{-0.15}$& 0.50$\pm$ 0.07&1.8\\
$\log (F_M/F_{Ln})$& S1& 23& 2& 0.18$^{+0.02}_{-0.05}$& 0.18$\pm$ 0.03&1.3\\
              & S2     & 16& 0& 0.27$^{+0.13}_{-0.06}$& 0.39$\pm$ 0.07&2.0\\
$\log (F_N/F_{Kn})$& S1& 38& 4& 0.87$^{+0.03}_{-0.04}$& 0.85$\pm$ 0.04&1.3\\
              & S2     & 16& 1& 1.69$^{+0.18}_{-0.53}$& 1.56$\pm$ 0.16&2.6\\
$\log (F_{Ln}/F_{Kn})$& S1& 30& 0& 0.36$^{+0.04}_{-0.02}$& 0.32$\pm$ 0.05&1.8\\
              & S2     & 11& 0& 0.75$^{+0.27}_{-0.10}$& 0.74$\pm$ 0.11&3.7\\
$\log (F_{Kn}/F_{Hn})$& S1& 57& 0& 0.27$^{+0.02}_{-0.01}$& 0.28$\pm$ 0.03&2.0\\
              & S2     & 22& 0& 0.19$^{+0.15}_{-0.02}$& 0.25$\pm$ 0.05&1.4\\
$\log (F_{Hn}/F_{Jn})$& S1& 57& 0& 0.20$^{+0.02}_{-0.0}$& 0.22$\pm$ 0.02&1.9\\
              & S2     & 22& 0& 0.23$^{+0.08}_{-0.02}$& 0.22$\pm$ 0.08&2.1\\
\hline
\end{tabular}

\end{center}
\vspace{3mm}
\end{minipage}
\normalsize

In  any case, the galaxies which  are known to have clear signs
of intense circumnuclear star formation (e.g., NGC 1068, NGC 1365, NGC
1808, NGC 3982, NGC 4388, NGC 4945, NGC 5427,  NGC 5728, NGC 5953, NGC
6221,  NGC 7469, NGC  7582, NGC 7592,  I 4553, Circinus galaxy), as is
evidenced mostly by spectroscopic investigations of nuclear spectra or
emission line imaging  surveys (e.g.  V\'eron-Cetty \&  V\'eron, 1986;
Pogge, 1989,  Evans  et  al., 1996),   do not  appear to   have colors
significantly different from the norm.\\

%%TAB2
\vspace{4mm}
\hspace{-4mm}
\begin{minipage}{9cm}
\renewcommand{\arraystretch}{1.2}
\renewcommand{\tabcolsep}{1.2mm}
\begin{center}
\vspace{-3mm}
TABLE 2\\
\vspace{2mm}
{\sc The p-values for the IR Color Distributions\\}
\footnotesize
\vspace{2mm}
\begin{tabular}{lcrrrrr}
\hline \hline

\multicolumn{1}{c}{Color}  & \multicolumn{1}{c}{Mean $p$} & \multicolumn{1}{c}{Median $p$} \\
\multicolumn{1}{c}{(1)}       & \multicolumn{1}{c}{(2)}    & \multicolumn{1}{c}{(3)}\\
\hline
$\log (F_{25}/F_{12})$& $<0.0001$ & $<0.0001$\\
$\log (F_Q/F_N)$      & $0.138$   & $0.087$  \\
$\log (F_N/F_{Ln})$   & $<0.0001$ & $<0.0001$\\
$\log (F_N/F_{M})$    & $0.059$   & $0.065$  \\
$\log (F_M/F_{Ln})$   & $0.007$   & $0.009$  \\
$\log (F_N/F_{Kn})$   & $0.002$   & $0.0001$ \\
$\log (F_{Ln}/F_{Kn})$& $0.0009$  & $0.0009$ \\
$\log (F_{Kn}/F_{Hn})$& $0.21$    & $0.21$   \\
$\log (F_{Hn}/F_{Jn})$& $0.63$    & $0.63$   \\
\hline
\end{tabular}

\end{center}
\vspace{3mm}
\end{minipage}
\normalsize

The color distributions are in general Gaussian and unimodal,
with some exceptions mentioned below. We have adopted the probability
$P_W$ associated with the W--test (Shapiro \& Wilk, 1965) in order to
estimate the Gaussianity of the color distributions for uncensored data. 

We have computed  the dispersions of the  color distributions for  the
uncensored data only. We have compared these observational dispersions
$\sigma_{obs}$ with the  values $\sigma_{cal}$ which are expected from
reasonable observational uncertainties  on the fluxes, using the first
order approximation  $\sigma(\log x)\simeq \log(e) \sigma(x)/x$. We have
adopted the  following percentage  errors:  10\% on $F_{25}$,  15\% on
$F_{12}$, 25\% on  $F_Q$,   20\% on  $F_N$,  20\%  on $F_M$, 15\%   on
$F_{Ln}$, 30\% on $F_{Kn}$, $F_{Hn}$, $F_{Jn}$. 

 As shown by  Table 3, the  expected values ($\sigma_{cal}$) are almost
always smaller than the observed ones ($\sigma_{obs}$).  This indicates
the presence of appreciable intrinsic dispersion in the IR colors. The
observed dispersions for S2s are typically greater  than those for S1s
even at the longest IR wavelengths,  where uncertainties in the fluxes
are similar  for the   two  types of objects.

Some comments  on the individual color distributions follow.

\subsubsection{The mid-IR colors}

Let us discuss the MIR colors $\log (F_{25}/F_{12})$ and $\log (F_Q/F_N)$.
The distribution of $\log (F_{25}/F_{12})$ for S1s presents a tail at
low values which makes its shape a little different from a Gaussian one
(the W-test applied to the uncensored data gives a probability $P_W$=0.026
that the distribution does not deviate from Gaussianity). For S1s, our
$\log (F_{25}/F_{12})$-distribution is shifted to somewhat lower values
than previous distributions based on much smaller data samples, whilst it
essentially confirms previous typical results for S2s (Giuricin et al.,
1995; Maiolino et al., 1995).

The $\log(F_Q/F_N)$-distribution, as yet unexplored in the literature, has
mean and median values similar to those of the $\log (F_{25}/F_{12})$--
distribution for S2s. On the other hand, S1s lead to a $\log
(F_Q/F_N)$-distribution which is shifted to somewhat greater values than
the $\log (F_{25}/F_{12})$ one.  \\

%%TAB3
\vspace{4mm}
\hspace{-4mm}
\begin{minipage}{9cm}
\renewcommand{\arraystretch}{1.2}
\renewcommand{\tabcolsep}{1.2mm}
\begin{center}
\vspace{-3mm}
TABLE 3\\
\vspace{2mm}
{\sc The Dispersions of the IR Color Distributions\\}
\footnotesize
\vspace{2mm}
\begin{tabular}{lccc}
\hline\hline
\multicolumn{1}{c}{Color}       &  
\multicolumn{1}{c}{$\sigma_{obs}$}         & 
\multicolumn{1}{c}{$\sigma_{obs}$}         & 
\multicolumn{1}{c}{$\sigma_{calc}$}
\\
\multicolumn{1}{c}{}       &  
\multicolumn{1}{c}{S1}         & 
\multicolumn{1}{c}{S2}         & 
\multicolumn{1}{c}{}
\\
\multicolumn{1}{c}{(1)}       & \multicolumn{1}{c}{(2)}    & 
\multicolumn{1}{c}{(3)}       & \multicolumn{1}{c}{(4)}\\ 
\hline
$\log (F_{25}/F_{12})$& 0.19& 0.21& 0.08\\
$\log (F_Q/F_N)$      & 0.12& 0.25& 0.14\\
$\log (F_N/F_{Ln})$   & 0.18& 0.40& 0.11\\
$\log (F_N/F_{M})$    & 0.17& 0.25& 0.12\\
$\log (F_M/F_{Ln})$   & 0.16& 0.27& 0.11\\
$\log (F_N/F_{Kn})$   & 0.23& 0.75& 0.16\\ 
$\log (F_{Ln}/F_{Kn})$& 0.27& 0.37& 0.15\\
$\log (F_{Kn}/F_{Hn})$& 0.19& 0.25& 0.18\\
$\log (F_{Hn}/F_{Jn})$& 0.13& 0.39& 0.18\\
\hline
\end{tabular}

\end{center}
\vspace{3mm}
\end{minipage}
\normalsize

The evaluation of the $\log
(F_{25}/F_{12})$ distributions for the 25 S1s and 17 S2 objects with
available $\log (F_Q/F_N)$ confirms the presence of some systematic
differences for S1s, since their distribution leads to a median value
of 0.31 and a mean value of 0.32. Consistently, a steepening of the MIR
SED of S1 objects towards shorter wavelengths is suggested by Roche et
al.'s (1991) MIR spectrophotometric observations. These authors noted that
the slopes of the $\lambda \sim$8--13 \m spectra they obtained on
ground--based telescopes with small beams ($\sim$ 5") were steeper than
the IRAS 12--25 \m slopes for galaxies with featureless MIR spectra
(mostly S1s). 

\subsubsection{The mid--to--near IR colors.}

The seeming bimodality of the $\log (F_N/F_M)$-histogram is not
statistically significant. The $\log (F_M/F_{Ln})$ distribution for S2s
presents a tail at high values, which makes it significantly non-Gaussian. 
($P_W$=0.002). The $\log (F_N/F_M)$ and $\log (F_M/F_{Ln})$ distributions
give results which are substantially consistent with those relative to the
$\log (F_N/F_{Ln})$ distributions, since they imply median (mean)
values of $\log (F_N/F_{Ln})$ equal to 0.63 (0.55) and 0.84 (0.89) for
S1s and S2 objects, respectively.  The $\log (F_N/F_{Ln})$-distribution
for S2s is significantly more peaked than a Gaussian distribution
($P_W<0.0001$). In the following we shall adopt the $\log (F_N/F_{Ln})$--
distribution in order to define the SED between the relevant wavelengths,
because the statistics are better and there is no significant indication of a
change (at $\lambda\sim$4.8 \m) in the SED slopes of the two types of
objects.  We point out that neglecting a correction for starlight in the
L-band, one obtains an appreciable overestimate of this color also for
small-beam observations of nearby objects. For instance, the $\log
(F_N/F_{L})$--distributions of 67 S1 and 66 S2 objects lead to
median (mean) values of 0.44 (0.42) and 0.74 (0.72), respectively.
Compared to the results of the corresponding
$\log(F_N/F_{Ln})$--distributions, this implies a median L-band starlight
contribution of 28\% and 55\% for the two groups of objects, respectively.
Correction for starlight is generally neglected in the relevant
literature, except in Pier \& Krolik (1993).  Encouragingly, relying on
the small data sample (15 S1s and 6 S2s) considered by these authors,
we obtain similar medians of $\log (F_N/F_{Ln})$ (i. e., $\sim$ 0.6
and 0.9, respectively). 

The two fluxes used to compute the quantities $\log (F_{25}/F_{12})$,
$\log (F_Q/F_N)$, $\log (F_N/F_M)$, $\log (F_M/F_{Ln})$, $\log
(F_N/F_{Ln})$ come from the same reference sources in many cases. This
limits the effect of data inhomogeneity on these five colors. This problem
may be more serious for the sample of $\log (F_N/F_{Kn})$, in which the
two fluxes generally come from different reference sources and refer to
measurements carried out in different epochs. We have obtained a $\log
(F_N/F_{Kn})$--distribution characterized by median (mean) values of 0.87
(0.87) for 51 S1s (with 6 upper limits on $F_N$) and 1.06 (1.30) for
38 S2s (with 4 upper limits on $F_N$). But we have preferred to rely
finally on a subsample of bona fide $\log (F_N/F_{Kn})$ colors, by
considering only objects with $F_{Kn}$ data taken from the accurate
decomposition analyses by Kotilainen et al. (1992a, b), Danese et al.
(1992), Zitelli et al. (1993), Kotilainen \& Prieto (1995), and 
Alonso--Herrero et al. (1996). Table 1 and Fig. 1 present our results for
this selected subsample, which, compared to the above-mentioned wider
sample, substantially confirms the median and mean values of the $\log
(F_N/F_{Kn})$ distribution for S1s, but leads to redder colors for
S2s. The addition of a few objects having $F_{Kn}$ estimates taken from Ward
et al. (1987) and McAlary \& Rieke (1988) yields a $\log (F_N/F_{Kn})$
distribution which is similar to that of the selected subsample. On
the other hand, the inclusion of $F_{Kn}$ fluxes taken from McLeod \&
Rieke (1995) and Heisler et al. (1996) tends to shift the $\log
(F_N/F_{Kn})$ distribution of S2s to appreciably smaller and larger
values, respectively, and gives rise to an even broader and more irregular
(i.e., non-Gaussian) color distribution than the broad one of the selected
subsample. The large proportion of $F_{Kn}$ data taken from McLeod \& Rieke
(1995) in the wide sample accounts for the above--mentioned difference in
the median colors of S2s. 

\subsubsection{The near-IR colors}

The sample of $\log (F_{Ln}/F_{Kn})$ colors leads to a distribution
characterized by median (mean) values of 0.34 (0.34) for 66 S1s (with
no censoring) and 0.65 (0.75) for 44 S2s (with one upper limit on
$F_{Ln}$). But this sample suffers from the same problems of inhomogeneity
which affect the $\log (F_N/F_{Kn})$ colors. Therefore, also in this case
we have preferred to rely on a selected subsample of bona fide $\log
(F_{Ln}/F_{Kn})$ colors, by considering only objects with both good
$F_{Kn}$ data (defined as above) and with $F_{Ln}$ fluxes derived from L
band data measured by the same authors in similar epochs; the latter
requirement is intended to avoid the possible effects of nuclear variability.
Table 1 and Fig. 1 refer to this selected subsample, which, however,
substantially confirms the typical values given by the wider sample for
both S1s and S2s. These two distributions were found to be not
significantly different from those of the selected subsamples through the
application of the usual two--sample tests, although the selected subset
of S2s seems to lead to a narrower spread of values and that of 
S1s shows a tail at low values ($P_W$=0.007). Incidentally, a
somewhat larger subsample which contains only objects with good $F_{Kn}$
data (as above), but with no restrictions on the reference sources for the
$F_{Ln}$ data, again essentially confirms the typical values 
listed in Table 1. 

The quantities $\log (F_N/F_{Ln})$- and $\log (F_{Ln}/F_{Kn})$ yield
results which are in substantial agreement with those given by the $\log
(F_N/F_{Kn})$-distribution, since for this quantity they imply
median values of 0.91 (for S1s) and 1.68 (for S2s).  The SEDs of both
types of objects considerably steepen beyond the L-band (i.e., at shorter 
wavelengths). In the following we shall use the colors $\log
(F_N/F_{Ln})$- and $\log (F_{Ln}/F_{Kn})$ to establish the SEDs between
relevant wavelengths.

As regards the quantities $\log (F_{Kn}/F_{Hn})$ and $\log
(F_{Hn}/F_{Jn})$, S2s show distributions which are broader than those of
S1s, but they are centered at similar values and have no significant
bimodality. We have verified that the distributions of theses two
quantities do not change appreciably, if we restrict these two
data samples to objects with $F_{Kn}$ data taken from the accurate
decomposition analyses by Kotilainen et al.  (1992a, b), Danese et al.
(1992), Zitelli et al. (1993), Kotilainen \& Prieto (1995), and 
Alonso--Herrero et al. (1996).\\

%%FIG2
\begin{minipage}{8.5cm}
\includegraphics{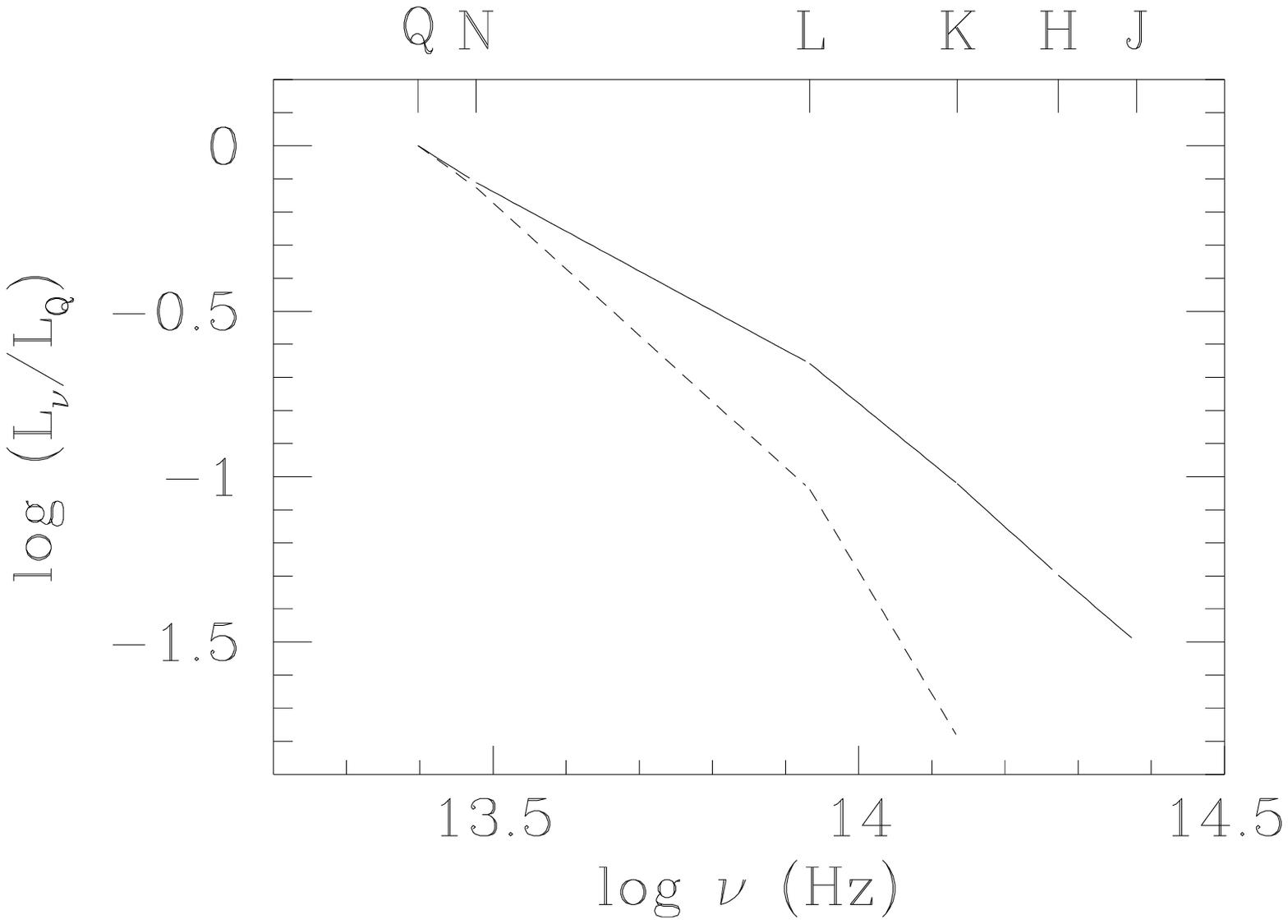}
$\ \ \ \ \ \ $\\
\vspace{6.5cm}
$\ \ \ $\\
\vspace{1mm}
{\small\parindent=3.5mm
{\sc Fig.}~2.---
This figure shows the median SEDs of S1 (solid
lines) and S2 nuclei (dashed lines). The two SEDs are arbitrarily
normalized so that they match at long wavelengths ($\lambda=20$ \m,
Q-band).
}
\vspace{5mm}

\end{minipage}

 The absence of any statistical difference
between S1s and S2s in these two colors is clearly unexpected. It is
likely to reflect very large uncertainties associated with the delicate
estimates of nuclear fluxes in the H and J bands for S2s, where the
starlight contribution is even greater than in the K band (see, e. g.,
Alonso--Herrero et al., 1996; Kotilainen et al., 1992a). Therefore, in the
following we will adopt take into consideration only the
results relative to S1s, disregarding those relative to S2s. \\

Fig. 2 illustrates the adopted median SEDs of S1 and S2 nuclei as
they result from the colors discussed above. The two SEDs are arbitrarily
normalized so that they match at long wavelengths ($\lambda$=20 \m). Since
S2s always have steeper SEDs than the former objects, the degree of
difference between the luminosities $L_{\nu}$ of the two groups of
objects increases towards shorter wavelengths.

\subsubsection{Correlations involving colors}

We have investigated the correlations between colors by computing the two
non-parametric rank correlation coefficients, Spearman's $r_s$ and
Kendall's $r_k$ (see, e.g., Kendall \& Stuart, 1977), in cases of
uncensored quantities. In testing the significance of the correlations
between the variables $x$ and $y$ containing censored data, we have relied
on the Cox proportional hazard model, the generalized Kendall rank
correlation statistics, and the generalized Spearman rank--order
correlation coefficient. The first method allows censored data for only
one variable, whereas the last two methods also allow simultaneous
censoring in both variables. 

As expected, the quantity $\log (F_{25}/F_{12})$ appears to
correlate  significantly with $\log (F_Q/F_N)$
for both S1s and S2s (at the median, one-tailed significant levels of
96.5\% and 99.2\%, respectively).\\

%TAB4
\vspace{4mm}
\hspace{-4mm}
\begin{minipage}{9cm}
\renewcommand{\arraystretch}{1.2}
\renewcommand{\tabcolsep}{1.2mm}
\begin{center}
\vspace{-3mm}
TABLE 4\\
\vspace{2mm}
{\sc The IR Luminosity Distributions\\}
\footnotesize
\vspace{2mm}
\begin{tabular}{lrrrrr}
\hline\hline
\multicolumn{1}{c}{Variable}       &  \multicolumn{1}{c}{Sample}       &  
\multicolumn{1}{c}{$N$}         & \multicolumn{1}{c}{$N_{ul}$}& 
\multicolumn{1}{c}{Median}      & \multicolumn{1}{c}{Mean}    
\\
\multicolumn{1}{c}{(1)}       & \multicolumn{1}{c}{(2)}    & 
\multicolumn{1}{c}{(3)}       & \multicolumn{1}{c}{(4)}    & 
\multicolumn{1}{c}{(5)}       &\multicolumn{1}{c}{(6)}\\ 
\hline
$ \log L_{25}$ & S1& 203& 45& 23.66$^{+0.03}_{-0.08}$& 23.57$\pm$ 0.07\\
               & S2& 335& 60& 23.57$^{+0.05}_{-0.03}$& 23.53$\pm$ 0.04\\
$ \log L_{12}$ & S1& 203& 69& 23.32$^{+0.09}_{-0.09}$& 23.24$\pm$ 0.08\\
               & S2& 332&131& 23.03$^{+0.06}_{-0.05}$& 23.01$\pm$ 0.05\\
$ \log L_{Q}$  & S1&  25&  0& 23.78$^{+0.23}_{-0.12}$& 23.73$\pm$ 0.13\\
               & S2&  20&  3& 22.99$^{+0.23}_{-0.16}$& 23.13$\pm$ 0.17\\
$ \log L_{N}$  & S1&  82& 15& 23.29$^{+0.17}_{-0.07}$& 22.94$\pm$ 0.15\\
               & S2&  91& 14& 22.58$^{+0.16}_{-0.08}$& 22.58$\pm$ 0.08\\
$ \log L_{M}$  & S1&  21&  0& 22.91$^{+0.23}_{-0.02}$& 22.83$\pm$ 0.16\\
               & S2&  15&  0& 22.47$^{+0.19}_{-0.52}$& 22.44$\pm$ 0.26\\
$ \log L_{Ln}$ & S1&  99&  2& 22.71$^{+0.09}_{-0.08}$& 22.65$\pm$ 0.07\\
               & S2& 101&  8& 21.79$^{+0.10}_{-0.06}$& 21.84$\pm$ 0.08\\
$ \log L_{Kn}$ & S1&  58&  0& 22.39$^{+0.19}_{-0.05}$& 22.41$\pm$ 0.07\\
               & S2&  21&  0& 21.13$^{+0.21}_{-0.26}$& 21.09$\pm$ 0.17\\
\hline
\end{tabular}

\end{center}
\vspace{3mm}
\end{minipage}
\normalsize

 In general we have detected no other
significant correlations, with the exception of a $\log (F_{25}/F_{12})$--
$\log (F_N/F_{Ln})$ correlation for S1 galaxies (N=68, 99.0\% significance
level), which holds also for uncensored data. 

We have verified that no color correlates with the galaxy distances
(for both S1s and S2s). Subsamples of nearby objects give similar
color distributions. 

Moreover, we have searched for correlations between colors and some
indicators of the MIR luminosities such as $L_N$, $L_{12}$, $L_{25}$,
because MIR emission, being an approximate constant fraction of the
bolometric flux of Seyfert nuclei, is believed to provide the best
indication of the nuclear bolometric luminosity (Spinoglio \& Malkan,
1989; Spinoglio et al., 1995). In general, we have detected no
correlations, except for the color $\log (F_{25}/F_{12})$, which shows
a slight tendency to be somewhat greater in S1 and S2 objects of
greater MIR luminosities. This behavior is dissimilar from that shown
by $\log (F_{60}/F_{25})$ and $\log (F_{100}/F_{60})$, which tend to
decrease with increasing MIR luminosity in Seyfert galaxies. The
latter behavior could be ascribed to a decreasing relative
contribution of the cool emission from galaxy disks in
high--luminosity AGNs (Giuricin et al., 1995).

\subsection{The IR luminosities}

We are interested in establishing the differences between the typical
luminosities of S1s and S2s in different bands, because appreciable
differences are predicted by torus models (especially for very optically
thick configurations and/or at the shortest IR wavelengths). For the sake
of comparison with samples published in the literature, in Table 4 we list
the total number of objects N, the number of upper limits $N_{ul}$, the
mean and median values of $L_{\nu}$ together with associated 1$\sigma$
errors, for the two groups of objects and different spectral bands.

Table 4 reveals that S1s tend to be more luminous than S2s at the shortest
IR wavelengths, but this may be partially due to observational selection
effects which favor the observation of luminous objects at great
distances (S1s are typically more distant than S2s). In order to remove
this distance selection effect  in the comparison of groups of objects
lying at different typical distances, for each spectral band we have
considered the logarithmic relation \begin{equation} \log L_\nu=17.078+log
(F_{\nu, min})+2 log D \end{equation} which yields the minimum value of $log
L$ (with $L_\nu$ in W/Hz) corresponding to the minimum uncensored or
censored value of $F_\nu$ ($F_{\nu, min}$) for an object at distance $D$ (in
Mpc). Obviously, upper limits and the detections of the faintest objects
cluster near this relation in the $\log L_{\nu}-\log D$ diagram. In the
case of $L_{Kn}$, we have taken the subset of objects with $F_{Kn}$ data
taken from the accurate decomposition analyses by Kotilainen et al.
(1992a, b), Danese et al. (1992), Zitelli et al.  (1993), Kotilainen \&
Prieto (1995), and Alonso--Herrero et al. (1996).

 We have  used relation (1)  to evaluate  the {\it observed} minus
{\it calculated}  value of $\log L_\nu$,   i. e. $\Delta(\log L_\nu)$,
for  each object.   In  this   manner  we have obtained   $\Delta(\log
L_\nu)$-distributions that  are free from distance  selection effects.
Table 5  lists the total   number of objects   N, the number  of upper
limits $N_{ul}$,  the mean and  median  values of $\Delta(\log L_\nu)$
(together with associated  1 $\sigma$ errors), for  the two groups  of
objects and different spectral bands. Fig. 3 displays the $\Delta(\log
L_\nu)$-distributions for   the most interesting   cases. There  is no
significant      evidence   of  bimodality    in   the    $\Delta(\log
L_\nu)$-distributions.  The statistical significance of the difference
between  the distributions  of  the  two groups of  objects  has  been
evaluated  through    the application  of   the  above--mentioned five
two--sample tests.  In several  cases the non--random character of the
censoring pattern tends to  give asymmetrical distributions.  However,
since two--sample tests seek only to compare  two samples, they do not
require that the censoring patterns of the two samples be random.

Table 5 and Fig. 3 reveal that S2 galaxies have, on average, greater
$L_{25}$ luminosities than S1s by a median factor of$\sim$1.5 (at a
significance level corresponding to $p<$0.0001). The $L_Q$ distributions
give consistent results, although the difference is never statistically
significant, because of the relatively low number of objects. As we go
towards shorter wavelengths the difference between the typical
luminosities of the two classes first disappears (see, e.g., $L_{12}$ and
$L_N$) and then reappears in the opposite sense (see $L_{Ln}$ and
$L_{Kn}$). As a matter of fact, the $L_{Ln}$ luminosities of S2s are
significantly fainter than those of S1s by a factor 1.8 (with a median
$p$=0.0004). In other words, as we go from $L_{25}$ to $L_{Ln}$, the ratio
of the S2 luminosities over S1 luminosities typically decreases by a
factor of $\sim$2.7. The $L_{Kn}$ luminosities of S2s are significantly
fainter than those of S1s by a factor $\sim$3.5 (with a median
$p<$0.0001); the total sample of $L_{Kn}$ data gives $\sim$3.7.
Hence, in going from $L_{25}$ to $L_{Kn}$, S2s become typically fainter than
S1s by a factor $\sim$5.2 ($\sim$5.5 for the total sample). We have
verified that subsamples of nearby objects always give similar results for
all spectral bands.

%FIG3
\end{multicols}
\includegraphics{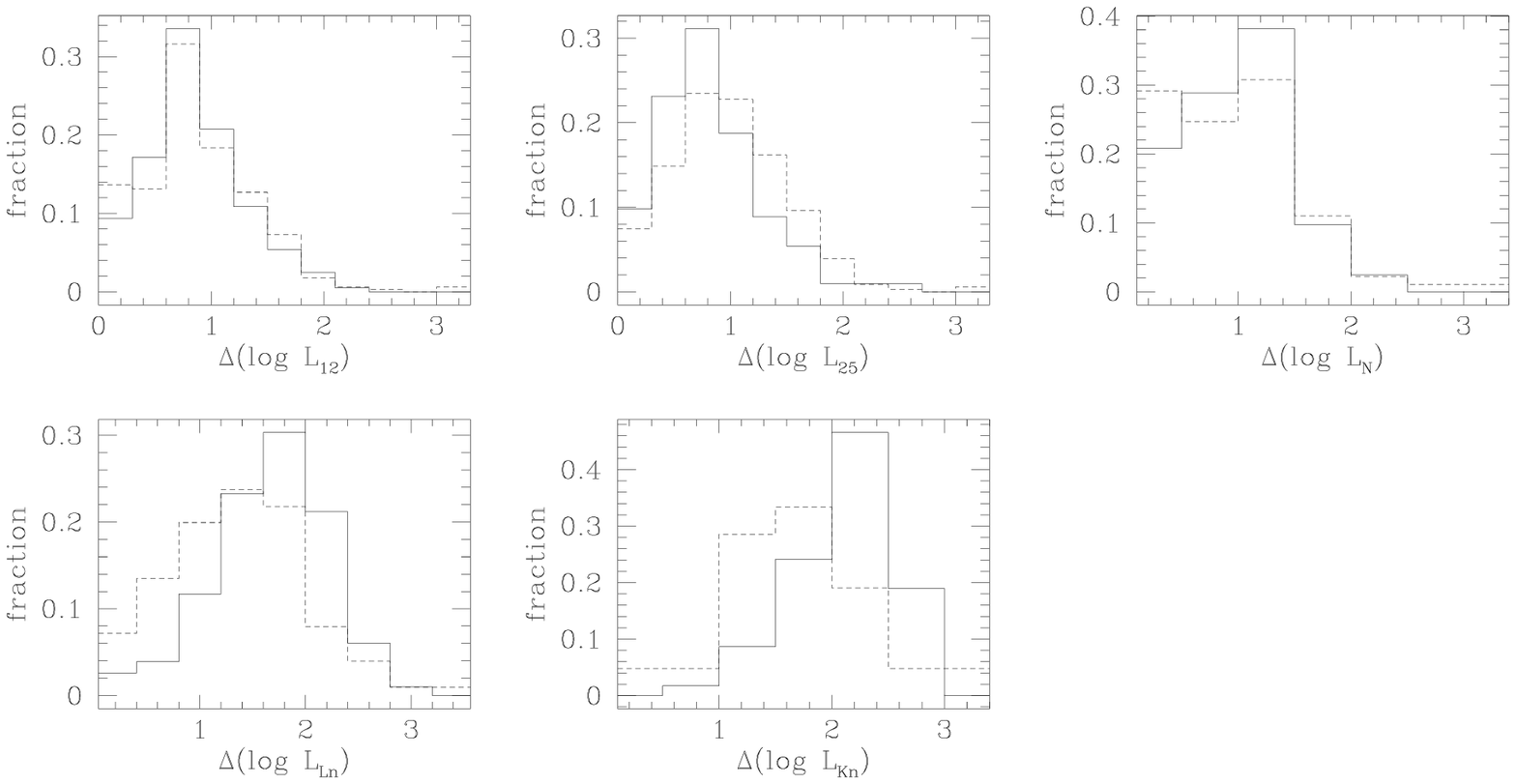}
$\ \ \ \ \ \ $\\
\vspace{9.cm}
$\ \ \ $\\
\vspace{1mm}
{\small\parindent=3.5mm
{\sc Fig.}~3.---
The histograms show the
distributions of $\Delta(\log L_{\nu})$ relative to different photometric
bands, for S1 (solid lines) and S2s (dashed lines).  $\Delta(\log
L_{\nu})$ is the {\it observed} minus {\it calculated} value of the
logarithm of the monochromatic luminosity.
}
\vspace{5mm}
\begin{multicols}{2}

%TAB5
\vspace{4mm}
\hspace{-4mm}
\begin{minipage}{9cm}
\renewcommand{\arraystretch}{1.2}
\renewcommand{\tabcolsep}{1.2mm}
\begin{center}
\vspace{-3mm}
TABLE 5\\
\vspace{2mm}
{\sc The $\Delta (\log L_{\nu})$ Luminosity Distributions\\}
\footnotesize
\vspace{2mm}
\begin{tabular}{lrrrrr}
\hline\hline
\multicolumn{1}{c}{Variable}       &  \multicolumn{1}{c}{Sample}       &  
\multicolumn{1}{c}{$N$}         & \multicolumn{1}{c}{$N_{ul}$}& 
\multicolumn{1}{c}{Median}      & \multicolumn{1}{c}{Mean}    
\\
\multicolumn{1}{c}{(1)}       & \multicolumn{1}{c}{(2)}    & 
\multicolumn{1}{c}{(3)}       & \multicolumn{1}{c}{(4)}    & 
\multicolumn{1}{c}{(5)}       &\multicolumn{1}{c}{(6)}\\       
\hline
$\Delta \log L_{25}$ & S1& 203& 45& 0.78$^{+0.04}_{-0.04}$& 0.81$\pm$ 0.04\\
                     & S2& 335& 60& 0.95$^{+0.03}_{-0.03}$& 1.01$\pm$ 0.03\\
$\Delta \log L_{12}$ & S1& 203& 69& 0.82$^{+0.03}_{-0.03}$& 0.86$\pm$ 0.04\\
                     & S2& 332&131& 0.82$^{+0.03}_{-0.03}$& 0.87$\pm$ 0.04\\
$\Delta \log L_{Q}$  & S1&  25&  0& 0.50$^{+0.21}_{-0.05}$& 0.61$\pm$ 0.07\\
                     & S2&  20&  3& 0.77$^{+0.21}_{-0.09}$& 0.85$\pm$ 0.14\\
$\Delta \log L_{N}$  & S1&  82& 15& 0.98$^{+0.13}_{-0.09}$& 0.97$\pm$ 0.06\\
                     & S2&  91& 14& 0.90$^{+0.11}_{-0.07}$& 0.96$\pm$ 0.07\\
$\Delta \log L_{M}$  & S1&  21&  0& 0.58$^{+0.02}_{-0.03}$& 0.60$\pm$ 0.08\\
                     & S2&  15&  0& 0.73$^{+0.10}_{-0.15}$& 0.82$\pm$ 0.15\\
$\Delta \log L_{Ln}$ & S1&  99&  2& 1.66$^{+0.07}_{-0.18}$& 1.66$\pm$ 0.06\\
                     & S2& 101&  8& 1.40$^{+0.08}_{-0.10}$& 1.37$\pm$ 0.07\\
$\Delta \log L_{Kn}$ & S1&  58&  0& 2.17$^{+0.09}_{-0.10}$& 2.10$\pm$ 0.06\\
                     & S2&  21&  0& 1.62$^{+0.11}_{-0.08}$& 1.64$\pm$ 0.16\\
\hline
\end{tabular}

\end{center}
\vspace{3mm}
\end{minipage}
\normalsize

Interestingly, the increasing dimming of S2s towards shorter wavelengths
(with respect to S1s) agrees with the differences between the SEDs of 
S1s and S2s and with the expectations of models of obscuring tori
emitting anisotropic radiation at short enough wavelengths. But it is a
little surprising that S2s appear to be typically brighter than S1s at
$\lambda\sim$25 \m, because in the framework of torus models S2s 
are expected to have similar or smaller luminosities. By the way,
in the same manner we have found that S2s are typically brighter than
S1s also at IRAS $\lambda\sim$60 and $\sim$100 \m bands by similar median
factors, i.e $\sim$1.5 and $\sim$1.7, respectively. 

We have deemed it useful to check this point by considering some subsamples,
namely the UV excess--selected Markarian objects, the IRAS 12 \m
flux--limited Seyfert galaxies listed by Rush et al. (1993), and the IRAS
sample by de Grijp et al.  (1992), who selected objects with relatively
warm 25 \m to 60 \m colors. Only the first and second subsamples confirm
the dissimilarity between the $\Delta(\log L_\nu)$-distributions of S1s
and S2s in the three above-mentioned IRAS bands. This dissimilarity turns
out to be fairly strong in the first subsample (where it amounts to a
factor of $\sim$1.6 for $\Delta(\log L_{25})$), weak in the second subsample
(where the corresponding factor is $\sim$1.3), whilst it disappears in the
third one. Hence, there is no sure evidence of an intrinsic difference
between the two classes of objects. Our results may simply reflect a
selection effect in several current samples, in the sense that S2s
undergoing active star formation are more likely to be detected at short
IR wavelengths and especially at optical wavelengths than objects not 
undergoing star formation. Thus star formation may seem more common in 
galaxy disks hosting S2 nuclei than in those containing S1s, 
as recently stressed by Maiolino et al. (1995), who examined
the extended MIR emission of Seyfert galaxies by comparing the
$\lambda\sim$10 \m ground--based observations with the IRAS
$\lambda\sim$12 \m fluxes. The aforementioned selection effect may also 
explain why circumnuclear star formation seems to be a little more 
common in S2s than in S1s, as suggested by the emission line imaging 
surveys of nearby Seyfert galaxies (e.g., Pogge (1989) and Evans et al. 
(1996)). 

We have provided a further check on this point by comparing the
distributions of the IR luminosities normalized to a quantity which is
powered by the active nucleus and is presumed to be isotropic, i.e free of
viewing angle effects, like optical emission lines produced in the NLR,
hard X-ray emission and central radio continuum emission (e.g., Mulchaey
et al., 1994). We have chosen to use the luminosity of the
[OIII]$\lambda$5007 A emission line, because it gives the widest data
sample, although this quantity can be affected by NLR dust absorption
(e.g., Dahari \& De Robertis, 1988a) and torus obscuration, as was
suggested in the case of radio galaxies (Jackson \& Brown (1990), Hes,
Barthel \& Fosbury (1993), Laing et al. (1994)). Although extended line
emission may be missed in small--aperture measurements, these measures are
less dependent than radio fluxes on the size of the observed region (the
central radio emission of Seyfert galaxies has been recently discussed by
Giuricin, Fadda \& Mezzetti, 1996). 

S1s and S2s have $\log(L_{[OIII]})$-distributions characterized by mean
(median) values of 41.15 (41.22) for N=140 objects and 40.90 (41.02) for
N=224 objects (with $L_{[OIII]}$ in erg/s), together with $\Delta\log
(L_{[OIII]})$-distributions characterized by corresponding values of 1.55
(1.57) and 1.56 (1.52). The two $\Delta\log (L_{[OIII]})$-distributions do
not differ, which is in line with the common view that the two types of
Seyfert galaxies have the same [OIII] properties (e.g., Dahari \& De
Robertis, 1988a; Keel et al., 1994; Mulchaey et al., 1994; Mulchaey,
Wilson \& Tsvetanov, 1996b).  Comparing the distributions of the quantities
$\log (L_{12}/L_{[OIII]})$, $\log (L_Q/L_{[OIII]})$, $\log
(L_{25}/L_{[OIII]})$, $\log (L_{60}/L_{[OIII]})$, $\log
(L_{100}/L_{[OIII]})$ by means of the usual two-sample tests we have 
again found that S2s have brighter normalized luminosities than S1s, in 
the last five bands, by similar median factors (i.e. $\sim$1.5, $\sim$1.4,
$\sim$1.8, $\sim$1.7, respectively), whereas they do not differ from S1s
at $\lambda\sim$12 \m. The analysis of the quantities $\log
(L_N/L_{[OIII]})$, $\log (L_{Ln}/L_{[OIII]})$, $\log (L_{Kn}/L_{[OIII]})$
shows that S2s have fainter normalized luminosities than S1s by median
factors of $\sim$1.5, $\sim$2.4, $\sim$5.9, respectively. Again we have
verified that subsamples of nearby objects give similar results. 

Incidentally, within extended data samples (though smaller than ours)
examined in the literature, de Grijp et al. (1992) and Keel et al. (1994)
found similar FIR luminosities for the S1 and S2 objects of their
IR--selected sample, whereas Dahari \& De Robertis (1988) and Mulchaey et
al. (1994) noted enhanced normalized IRAS luminosities (at least in one of
the above--mentioned three bands) for the S2s of their samples which are
dominated by Markarian objects. All these results are consistent with what
we found for the overall sample and subsamples based on different 
selection criteria. \\

%FIG4
\begin{minipage}{8.5cm}
\includegraphics{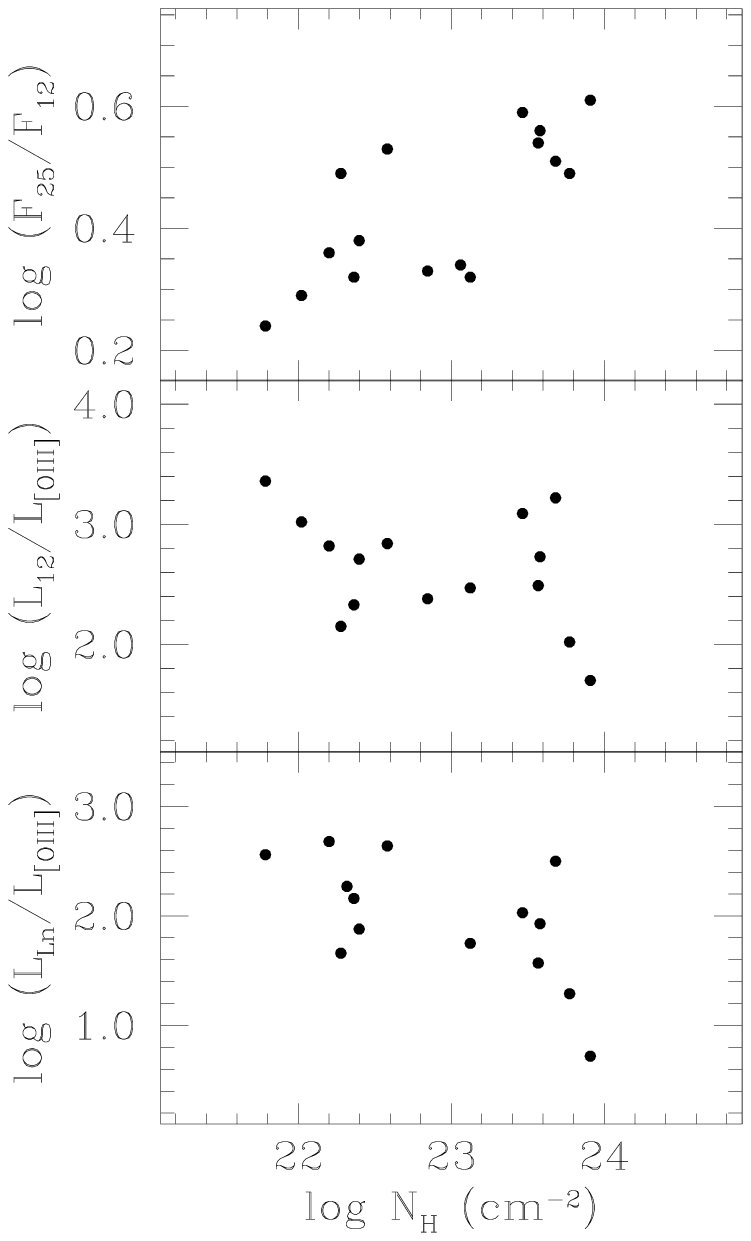}
$\ \ \ \ \ \ $\\
\vspace{11cm}
$\ \ \ $\\
\vspace{1mm}
{\small\parindent=3.5mm
{\sc Fig.}~4.---
The plots show the correlations between infrared
quantities and the absorbing hydrogen colums $N_H$ (expressed in
$cm^{-2}$). The infrared luminosities are integrated luminosities
evaluated as $\nu$ $L_{\nu}$.
}
\vspace{5mm}

\end{minipage}

In any case, our results clearly favor models in which obscuring
structures are essentially isotropic emitters already at $\lambda\sim$25
m. It is encouraging that the degree of the increasing faintness of S2s
towards shorter wavelengths (with respect to S1s) -- as derived from a 
comparison of the luminosities -- turns out to be in fairly good,
quantitative agreement with what we expect from the differences between
the SEDs of the two groups of objects. S2s are found to be fainter than
S1s by a typical factor of $\sim$1.5 at $\lambda\sim$12 \m, of $\sim$2.7
at $\lambda\sim$3.5 \m, of $\sim$5-7 at $\lambda\sim$2.2 \m. (Specifically
in the last case, values of $\sim$5.2 and $\sim$6.6 are derived from the
differences in the luminosities and in the SEDs, respectively.) The
analysis of the IR luminosities normalized to $L_{[OIII]}$ -- an analysis
which we regard as less reliable than the previous ones -- gives similar
results at long IR wavelengths and gives a somewhat more pronounced
dimming of S2s at short IR wavelengths. 

Within the framework of torus models these numbers establish the degree of
anisotropy of the IR emission. Notably, we find a lower degree of
anisotropy than that predicted by many torus models (see next section)
and claimed by Heckman (1995). For instance, examining the distributions
of $\log (L_N/L_R)$, where $L_R$ is the central radio luminosity at the
frequency of 1.4 GHz), Heckman (1995) argued that S1s are typically
brighter than S2s by a factor\\

%TAB6
\vspace{4mm}
\hspace{-4mm}
\begin{minipage}{9cm}
\renewcommand{\arraystretch}{1.2}
\renewcommand{\tabcolsep}{1.2mm}
\begin{center}
\vspace{-3mm}
TABLE 6\\
\vspace{2mm}
{\sc Correlations between N$_H$ and IR Properties\\}
\footnotesize
\vspace{2mm}
\begin{tabular}{lrrr}
\hline\hline
\multicolumn{1}{c}{Variables}       &  
\multicolumn{1}{c}{N}         & 
\multicolumn{1}{c}{r$_S$}         & 
\multicolumn{1}{c}{r$_K$}
\\
\multicolumn{1}{c}{(1)}       & \multicolumn{1}{c}{(2)}    & 
\multicolumn{1}{c}{(3)}       & \multicolumn{1}{c}{(4)}\\    
\hline
$N_H - \log (F_{25}/F_{12})$    & 16 & 0.71 (99.90\%) &  0.52 (99.76\%) \\
$N_H - \log (L_{12}/L_{[OIII]})$& 15 &$-$0.36 (90.44\%) & $-$0.22 (87.25\%) \\
$N_H - \log (L_{Ln}/L_{[OIII]})$& 14 &$-$0.58 (98.56\%) & $-$0.45 (98.76\%) \\
\hline
\end{tabular}

\end{center}
\vspace{3mm}
\end{minipage}
\normalsize

 of $\sim$4 (a result which is plagued by the
large inhomogeneity of the radio fluxes); the distributions of $\log
(L_N/L_{[OIII]})$ yielded a lower factor of $\sim$2, which is less
inconsistent with ours. Our results are closer to those of Maiolino et al. 
(1995) and Giuricin et al. (1995), who derived a milder anisotropy from
the direct comparisons of the $L_N$ luminosities. 

\subsection{Relation between the IR properties and X-ray absorption}

As is expected within the framework of torus models, the NIR and MIR
emissions of Seyfert nuclei appear to be related to their hard X-ray
emission (see Danese et al. (1992), Kotilainen et al.  (1992b) and
Giuricin et al. (1995), Barcons et al. (1995), respectively). 
Furthermore, Goodrich et al. (1994) noted that the S2 galaxies in which
broad Pa$\beta\lambda1.2818$\m emission lines have been detected are also
the objects which generally have the lowest values of the X-ray absorbing
hydrogen columns $N_H$ derived from the hard X-ray spectra. Recently,
Granato et al. (1997) found some evidence that the X-ray column depths of
S2 galaxies show a negative correlation with their N-band emissions, at
least for objects which have a so low Compton depth ($N_{H}\lesssim10^{24}
cm^{-2}$) that the hard X-rays can be seen directly through the torus. 

Our wide sample of IR data allows us to investigate further this point. 
To this end we have assembled the $N_H$-values (mean values in the
case of multiple entries) as derived by Smith \& Done (1996) (for
power-law fits to GINGA data), Iwasawa (1996), Warwick et al. (1993) for
17 Compton-thin S2 objects. Their $N_H$--distribution ranges over $6\cdot 
10^{21}\leq N_H\leq 10^{23}$ (with $N_H$ in units of $cm^{-2}$), with a
median value of $N_H=7\cdot 10^{22}$, which is considerably greater than the
median value $N_H\sim10^{21}$ of S1 objects (Nandra \& Pounds, 1994), 
even though the S2 sample is biased towards low values of $N_H$. 

We examined the correlations between $N_H$ and the IR colors discussed in
\S 3.1 and those between $N_H$ and the IR luminosities (normalized to the
$L_{[OIII]}$ luminosities) discussed in \S 3.2. As regards the colors,
only $\log (F_{25}/F_{12})$ displays a significant (positive) correlation
with $N_H$; the other colors show no correlations (but the number of
objects is smaller). As regards the luminosities, $\log
(L_{12}/L_{[OIII]})$ shows a marginal anti-correlation with $N_H$ (but
$\log (L_{25}/L_{[OIII]})$ does not for the same objects) and $\log
(L_{Ln}/L_{[OIII]})$ displays a strong anti-correlation; the lack of
correlation for the other colors may simply be due to the smaller number
of objects. Fig. 4 illustrates the three above-mentioned significant
correlations and Table 6 presents the relevant results. In this table we
list the two correlation coefficients, Spearman's $r_s$ and Kendall's
$r_k$, together with the associated (one-tailed) percent significant
levels (in these cases we have only detected objects). 

If we added in also the data for a few S2 galaxies which are likely to
have Compton-thick ($N_H>10^{24} cm^{-2}$) obscuring structures (i. e.,
NGC 1068, NGC 4945, NGC 6552, and the Circinus galaxy; see, e.g., Matt et
al., 1996, and references cited therein) so that their hard X-ray emission
is presumably due to a scattered nuclear component, all the correlations
would vanish. This enlarged, less biased, sample of S2 objects has a
median value of $N_H\sim2.9\cdot 10^{23} cm^{-2}$. 

In summary, our study provides further support for the contention  
that the IR emitting torus is related to the X-ray absorbing
structure, but only for Compton-thin structures. 

\section{Comparison with theoretical models}

The IR colors and the degree of anisotropy of the IR emission derived in
the previous section can be compared with the relevant predictions of a
wide series of torus models (PK, GD, ER) characterized by different
geometries, opacities and viewing angles. We have computed the theoretical
colors and anisotropies by filtering the predicted IR SEDs through the
typical overall spectral responses (including filter transmission,
detector response and atmospheric transmission), which were taken from
Bessell \& Brett (1988) for the JHKL photometry, from Campins, Rieke \&
Lebofsky (1995) for the NQ photometry, from Beichman et al. (1988) for
IRAS photometry. 

\subsection{Models of compact cylinders}
 
PK modeled the torus simply as a geometrically thick, annular ring of
uniform density dust which surrounds a central point source of radiation. 
PR proposed cylindrical configurations characterized by small extension
(radial sizes $\lesssim$ a few pc) and very large optical thickness, in
order to fit the general shape of the IR SED of Seyfert galaxies (in
particular that of the S2 galaxy NGC 1068) and the appearance of the
$\sim$ 10 \m silicate feature, which is seen in absorption in many S2
objects, whereas it is undetected in many S1 galaxies. Specifically, they
preferred models with radial and vertical Thomson depths
$\tau_r$=$\tau_z$=1 (corresponding to a visual extinction $A_V\sim$750 mag
and a hydrogen column density $N_H=1.5\cdot 10^{24} {\rm cm}^{-2}$ for
standard dust and dust-to-gas ratio), an inner radius-to-height ratio
$a/h\sim0.3$, and $T_{\rm eff}\sim$500-1000 K (where $T_{\rm eff}$ is the
effective temperature of the illuminating radiation on the inner wall of
the torus).  Basically, models with greater values of $a/h$ or $T_{eff}$
would shift the peak of the thermal bump to too short wavelengths and
would lead to too warm IR colors, whereas models of low optical thickness
($\tau_r$=0.1 and/or $\tau_z$=0.1) would predict an (unobserved) prominent
10 \m emission feature for face-on viewing angles. 

Examining the predicted colors for face-on views (viewing angle
$i=0\fdg$), we have noticed that published SEDs (for
$T_{eff}\sim$1000--2000 K) are in general characterized by too warm
(small) IR colors, with SEDs which peak at shorter wavelengths than
observed. Only models with low values of $T_{eff}$ ($T_{eff}\sim$500 K)
and $a/h$ ($a/h\sim$0.1) can reasonably fit the MIR colors (i.e. $\log
(F_{25}/F_{12})$ and $\log (F_Q/F_N)$) of S1 objects, for the wide range
of values of Thomson optical depths ($\tau\sim$0.1-10) explored by PK. But
at shorter IR wavelengths those models imply a too steep IR SED, even for
the choice of $\tau$-values which minimizes this problem (i. e.,
$\tau_z=\tau_r=1$); the relevant colors would require models with greater
values of $a/h$ and $T_{eff}$ ($a/h\sim$0.3-1, $T_{eff}\sim$1000 K). 
Similar problems hold for explaining both the long and short
wavelength sides of the SEDs of S2s at the same time.

Recognizing a problem which is emphasized by our wide data sample, PK
noted that the $\log (F_N/F_{Ln})$ colors predicted by their preferred
models were appreciably greater than the observed colors of a small sample
of 15 S1 galaxies. In order to make the observed mid-to-near IR
slope of these galaxies and PG quasars fit better with a broader IR bump, PR
introduced a black body emission component ascribed to a small amount of
hot ($T\sim$1300 K) dust which lies inside the inner edge of the torus
and is hidden by its main body from edge-on views.

But the preferred optically thick tori, with $\tau_z=\tau_r=1$, $a/h=0.3$,
$T_{eff}\sim500-1000$ K, entail an unreasonably large degree of anisotropy
in the IR emission. For instance, they predict a ratio between the
luminosities received by a polar ($i=0\fdg$) and equatorial ($i=90\fdg$)
observer $L_{\nu}(p)/L_{\nu}(e)$ equal to $\sim$4-7, $\sim$10-20,
$\sim$300-1500 at $\lambda\sim$25, \m, $\sim$12 \m, $\sim$3.5 \m,
respectively. For $\cos i=0.25$ in place of $\cos i=0$, the
corresponding ratios decrease respectively to $\sim$2.5-3.0, $\sim$3-5,
$\sim$7-10, which are still large values.  Only tori of low opacity, which
however do not fit other characteristics, would predict a mild enough
degree of anisotropy; for instance, tori with $\tau_z=\tau_r=0.1$,
$a/h$=0.3, $T_{eff}=1000$ K) imply $L_{\nu}(p)/L_{\nu}(e)$ equal to
$\sim$1.7 and $\sim$2.5 at $\lambda\sim$ 25 \m and $\sim$12 \m. 
   
\subsection{Models of tapered disks}

In later models (GD, ER) inhomogeneous configurations with different outer
shapes and density gradients were explored.  ER discussed three different
types of extended models, i. e. flared discs, anisotropic spheres, and
tapered discs. They soon realized that flared discs suffer the main
drawback of predicting very prominent 10 \m emission features for face-on
views. We have found that the two published models give too small colors
for face-on views. For edge-on views, opaque models with equatorial
optical depth $\tau_{uv}=200$ (where $\tau_{uv}=\tau_{\lambda=1000 \\A}$)
reproduce colors better than configurations with relatively low opacity
($\tau_{uv}$=40), except for $\log (F_N/F_{Ln})$, which still remains too
low, but they predict too large an anisotropy of the IR emission (for
standard dust and dust-to-gas ratio $A_V\sim$0.24 $\tau_{uv}$). The authors
noted that very optically thick and compact flared disks would reduce the
strength of the emission feature, but they would also produce too narrow 
an IR continuum. 
 
Anisotropic spheres yield SEDs which show a more gradual transition from
an edge-on to a face-on view than flared disks do. But face-on views still
display strong 10 \m emission features and SEDs that are too flat in the MIR
interval, whereas edge-on views, though predicting fairly realistic colors
for opaque configurations ($\tau_{uv}=200$), with the exception again of
$\log(F_N/F_{Ln})$ which remains too low, give too large an anisotropy. 
 
The authors devised tapered discs (discs whose height increases with
distance from the central source but tapers off to a constant height in
their outer part) mainly to suppress the 10 \m emission silicate features
from face-on views, without imposing an excessive equatorial depth (which
would entail an unreasonably narrow IR continuum) or without resorting to
lower silicate efficiencies (as proposed by Laor \& Draine, 1993) or
silicate depletion by shocks (as proposed by GD). The tapered disk models
assume the following main parameters: the opening angle of the disk
$\Theta_1$, the ratio of inner to outer radii $r_1/r_2$, the ratio of the
maximum half-thickness of the disc to the outer radius $h/r_2$, the
equatorial optical depth $\tau_{uv}$, and the dust density distribution in
radius. 

The authors asserted that configurations with large opening angles
($\Theta_1\sim60\fdg$) of the disk are unrealistic because they tend to
show 10 \m absorption features from all viewing angles, whereas
configurations with small opening angles ($\Theta_1\sim30\fdg$) show
implausible emission features for face-on views.  Therefore, the authors
deemed tapered disks with intermediate opening angles to be realistic and
published the SEDs of four models with $\Theta_1=45\fdg$, among which they
preferred the one which predicts an almost featureless MIR continuum
together with a broad IR SED for face-on views. The preferred model is
characterized by the following parameters:$\Theta_1=45\fdg$, 
$\tau_{uv}=1500$, $r^{-1}$ density distribution, $r_1/r_2$=0.01,
$h/r_2=0.3$.

We have found that none of the published models of tapered discs
satisfactorily reproduce the observed small degree of anisotropy and the
observed IR SEDs for face-on or edge-on views. In particular, the
above-mentioned preferred model exhibits colors too large at longer IR
wavelengths and too small at shorter IR wavelengths for edge-on views,
together with too large colors at shorter IR wavelengths for face-on
views. Besides, the model implies a large anisotropy corresponding to 
$L_{\nu}(p)/L_{\nu}(e)$ equal to $\sim$10 and $\gtrsim$50 at $\lambda\sim$25 
\m and $\sim$12 \m. For a viewing angle of $i=60\fdg$ the  
anisotropy, which remains large, decreases to $\sim$2.5 and $\sim$7 at
$\lambda\sim$25 \m and $\sim$12 \m, respectively.
  
In conclusion, models of tapered disks, though having the virtue of
solving problems related to the appearance of the 10 \m feature, do not
seem to easily reproduce the typical shape of the NIR and MIR SEDs of
Seyfert nuclei and the observed anisotropy.  

Inspecting in detail NGC 1068, Efstathiou, Hough \& Young (1996) realized
that tapered disk models alone were unable to describe satisfactorily the
IR SED of this object. \\

%FIG5
\begin{minipage}{8.5cm}
\includegraphics{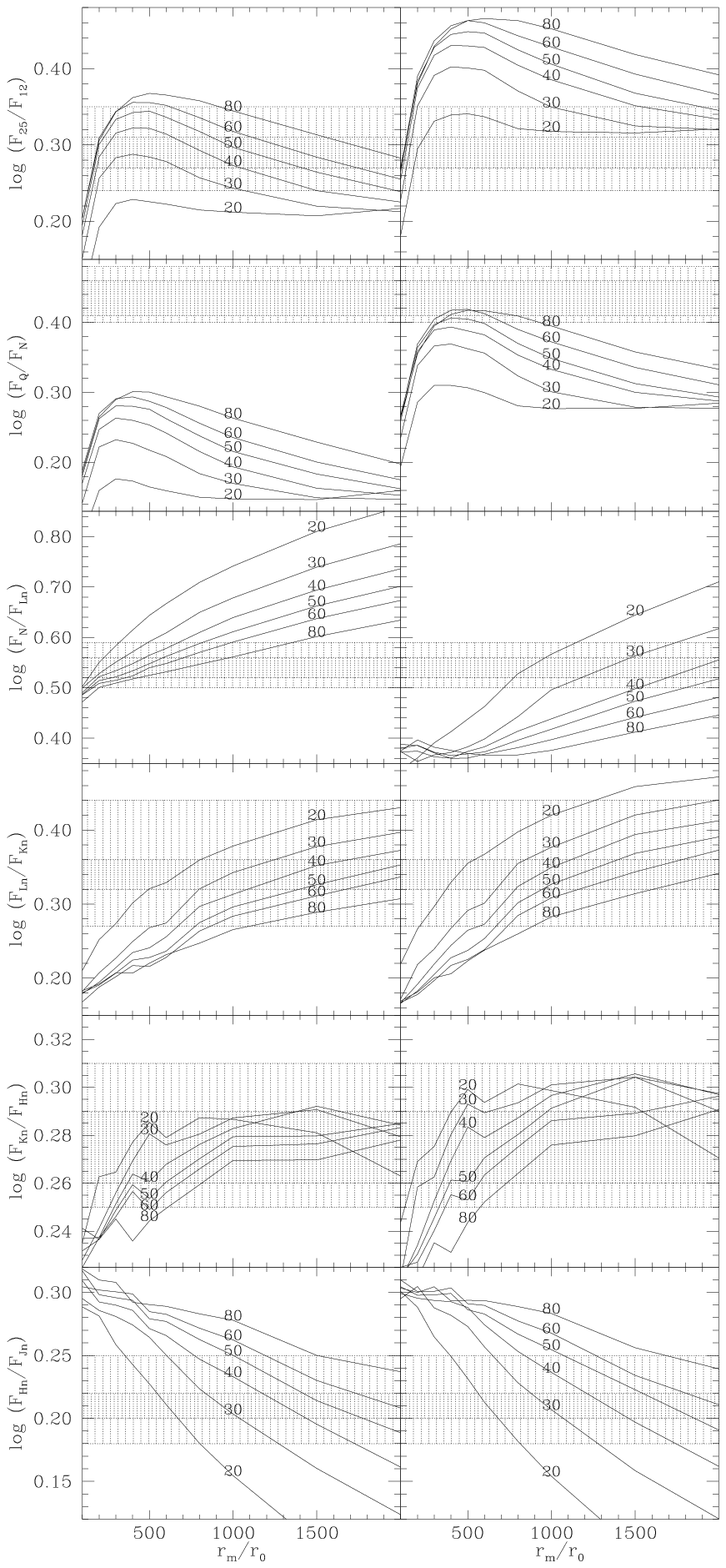}
$\ \ \ \ \ \ $\\
\vspace{19.5cm}
$\ \ \ $\\
\vspace{1mm}
{\small\parindent=3.5mm
{\sc Fig.}~5.---
The plots show the colors predicted by torus models
as a function of the outer--to--inner radius $r_m/r_0$, for values the
equatorial optical thickness $\tau_e$ ranging from 20 to 80, in case of
a face-on viewing angle. The two hatched regions indicate the $1-\sigma$ and
$2-\sigma$ error bars associated with the observed median. The left and
right columns refer respectively to models with a standard silicate
abundance and with a depression of silicates in the innermost
regions.
}
\vspace{5mm}
\end{minipage}

Therefore, they supplemented the emission from the
torus with an NIR component which was attributed to optically thin dust
distributed in the ionization cone. In this way they obtained a more
realistic SED, although the detailed fit of the 10 \m feature and of the
extension of the NIR and MIR emissions are still difficult problems.  We
note that their composite model still implies a largely different SED for
objects which are seen face-on and edge-on. 

\subsection{Models of flared disks}

GD proposed flared disk configurations characterized by fairly large
radial extension (from tens to hundreds pc), moderate optical thickness,
an almost homogeneous density distribution, i.e. tori with equatorial
optical depth $\tau_e=\tau_{\lambda 3000 \\A}\sim10-100$ (for standard
dust and dust-to-gas ratio $A_V\sim 0.61\cdot \tau_e$ mag. and
$N_H\sim1.2\cdot 10^{21} \tau_e cm^{-2}$), density distribution
$\rho\propto r^{-\beta}$ with $\beta\le0.5$, outer-to-inner radius ratio
$r_m/r_0 \sim 300-1500$ (where $r_0=0.5\ (L/10^{46}{\rm erg/s})^{1/2}$ pc
is set by the condition of sublimation of all dust species). In this way
GD were able to reproduce the average shape of the IR SEDs of a sample of
16 optically-selected S1 objects, without resorting to additional IR
emission components. In order to suppress the occurrence of the 10 \m
silicate emission in these models for face-on views, GD proposed that
shocks induced by radiation pressure depress the abundance of silicate
grains in the innermost regions of tori (at $r< \mbox{few} \cdot 10 r_0$). 

Recently, from several arguments, which include the $\log F_N/F_{Ln}$
colors of a few nearby S2 galaxies and an upper limit on anisotropy in the
MIR emission, Granato et al. (1997) argued in favor of moderately thick
tori (with line-of-sight extinction $A_V\lesssim$80 mag) rather than of
very thick configurations. Moreover, these authors fitted the IR SED of
NGC 1068 with a line-of sight extinction $A_V\sim50-100$ mag and with
slightly less extended tori (with $r_m/r_0\sim 100-200$) than those
proposed for S1s. If this were a general property of S2 galaxies, it would
be a serious problem for the unified model of Seyfert galaxies. Clearly,
this point deserves a statistical investigation such as the one we 
are carrying out in this paper. 
 
Tori of moderate optical thickness such as those proposed by GD certainly
promise to be successful models for explaining our observational results,
which lead to a quite low degree of IR emission anisotropy. Thus, we have
decided to undertake a detailed comparison between a wide series of GD
torus models and observational data, in order to assess to what
extent these models are compatible with our (partially new) results, which
could provide more stringent constraints on torus properties. Using the GD
code, we have computed the IR SEDs of many torus models for which we have
always assumed a uniform density distribution ($\beta$=0) for a standard
dust composition and a covering factor $f$=0.8, in order to keep the
number of free model parameters to a minimum. We have parameterized this
class of models by the two parameters $r_m/r_0$ and $\tau_e$. The former
parameter is related mainly to the broadness of the IR bump, whereas the
latter controls the near-to-mid infrared slope of the SED along obscured
lines of sight and the degree of its anisotropy. 

Fig. 5 shows the colors predicted by the GD torus models as a function of
$r_m/r_0$ for different choices of the equatorial optical thickness
$\tau_e$, in cases of face-on viewing angles. Fig. 6 show the analogous 
plots in cases of edge-on viewing angles. In Figs. 5 and 6    
the two hatched regions mark the 1--$\sigma$ and 2--$\sigma$ error
bands associated with the observed median. We show models with a
standard silicate abundance and models with a typical depression of
silicate grains in the innermost regions (i.e., for $r\leq50 r_0$). The
latter models are intended to give an indication of the effects of a
silicate depression on the colors. 
 
%FIG6
\begin{minipage}{8.5cm}
\includegraphics{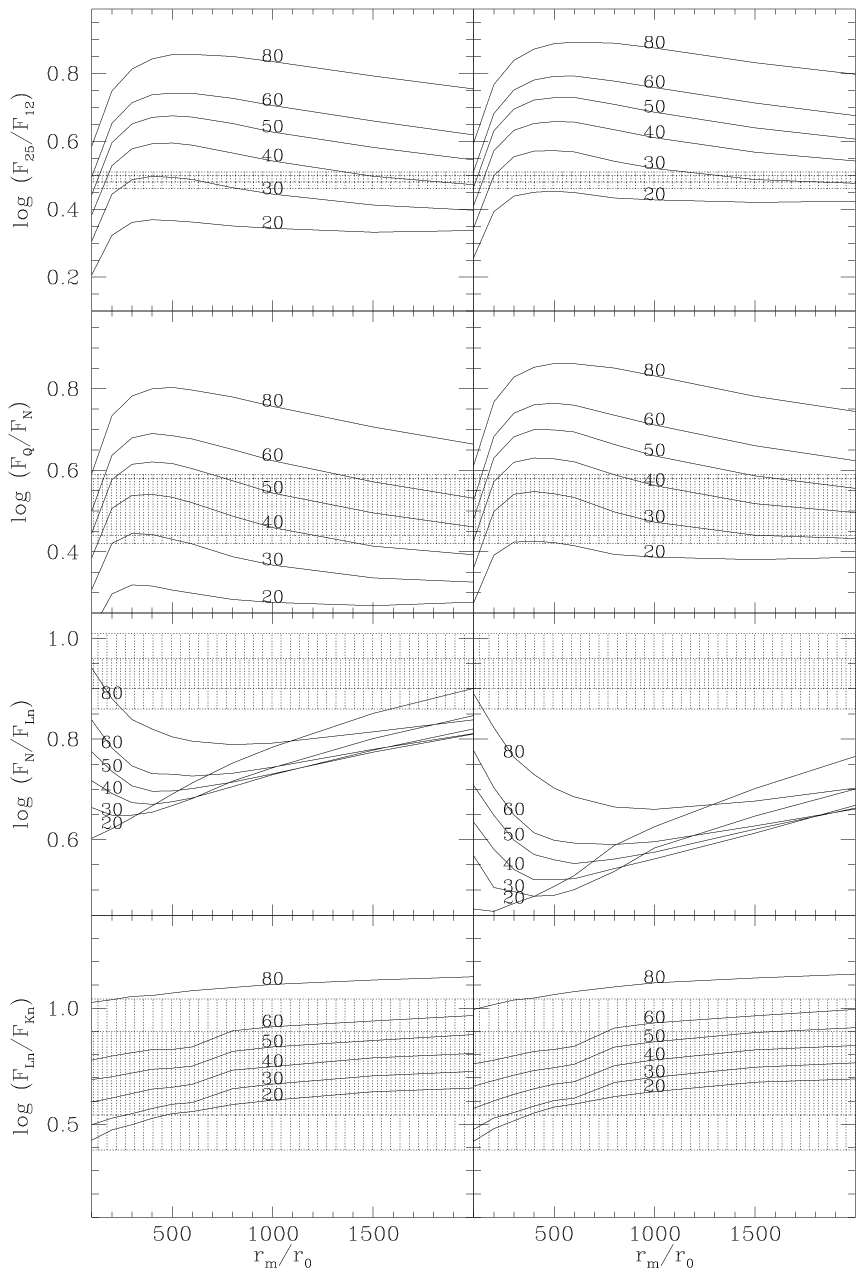}
$\ \ \ \ \ \ $\\
\vspace{13.5cm}
$\ \ \ $\\
\vspace{1mm}
{\small\parindent=3.5mm
{\sc Fig.}~6.---
Same as in Fig. 5, but for an edge-on viewing 
angle.
}
\vspace{5mm}
\end{minipage}

Let us first look at the models with a standard silicate abundance. They  
always tend to give values for the color $\log (F_{25}/F_{12})$ that are 
greater than for $\log (F_Q/F_N)$ rather than vice-versa, as observed.
The relatively high values of the latter quantity, namely the steepening
of the SED as we go towards shorter wavelengths within the MIR spectral
range, is not adequately reproduced by models. \\

%FIG7
\begin{minipage}{8.5cm}
\includegraphics{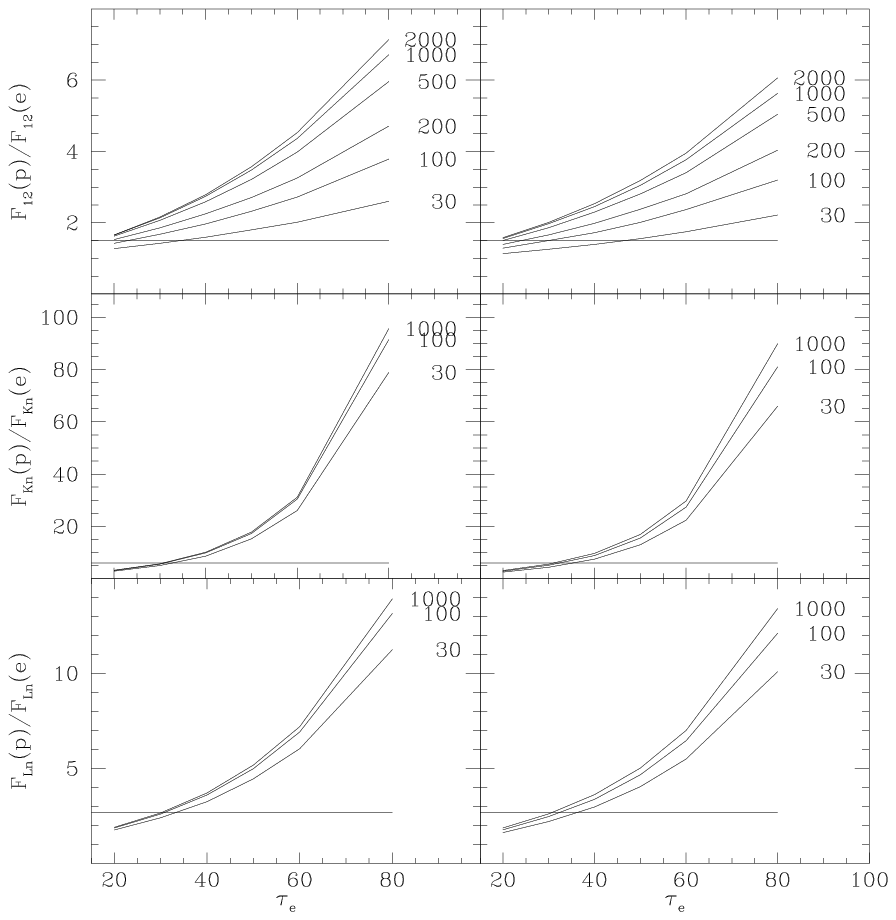}
$\ \ \ \ \ \ $\\
\vspace{10.cm}
$\ \ \ $\\
\vspace{1mm}
{\small\parindent=3.5mm
{\sc Fig.}~7.---
For some photometric bands the
plots show the ratio between the fluxes received by a polar and an
equatorial observer as a function of the equatorial optical thickness
$\tau_e$, for different values of the outer--to--inner radius $r_m/r_0$.
The horizontal lines denote the median values derived from the comparison
of the IR luminosities of S1 and S2 nuclei. The left and right columns
refer respectively to models with a standard silicate abundance and with a
depression of silicates.
}
\vspace{5mm}
\end{minipage}

It is thus reasonable to
be content with reproducing a typical mean MIR color $\log
(F_{25}/F_{12})\sim\log (F_Q/F_N)$ of $\sim$0.3-0.4. These values rule out
compact tori (i.e. with $r_m/r_0\lesssim100$). The
color $\log (F_N/F_{Ln})$ is easily accounted for by models characterized
by relatively low values of $r_m/r_0$ and $\tau_e$ (e.g., 
$r_m/r_0\sim$200 and
$\tau_e\sim20$) or by relatively high values of the two parameters (e.g.,
$r_m/r_0\sim$700 and $\tau_e\sim$80). In this case a decrease of $r_m/r_0$
can be counterbalanced by an adequate decrease of $\tau_e$. A similar
correlation between the two parameters holds for the color $\log
(F_{Ln}/F_{Kn})$, which rejects models with low $r_m/r_0$-values (e.g.,
with $r_m/r_0\lesssim$300 for $\tau_e\sim$20 or with $r_m/r_0\lesssim$1000
for $\tau_e\sim$80). The color $\log (F_{Kn}/F_{Hn})$ tends to reject
tori of relatively small size and for very extended tori is not very 
sensitive to changes of $r_m/r_0$ and $\tau_e$. The relatively low value
of $\log (F_{Hn}/F_{Jn})$ even more clearly favors very extended tori
($r_m/r_0\gtrsim500$), within the usual $\tau_e$-interval. 

The inclusion of silicate depression leaves the NIR SED almost unaffected,
increases the MIR colors, and considerably reduces the color
$\log (F_N/F_{Ln})$. These effects are similar for all viewing angles. 
For face-on viewing angles the MIR colors are now in better agreement with
the observations. Now also $\log (F_N/F_{Ln})$ agrees with the quantities
$\log (F_{Ln}/F_{Kn})$, $\log (F_{Kn}/F_{Hn})$, $\log (F_{Hn}/F_{Jn})$ in
favoring very extended configurations. In summary, the IR colors better
fit models of very extended tori($r_m/r_0\gtrsim$500);  $\tau_e$ is not
well constrained within the usual range ($\tau_e\sim$20-80).  Notably,
much larger and much lower values of $\tau_e$ would give too steep and too
flat MIR SEDs, respectively, for non-compact tori. 
 
Let us discuss edge-on systems, disregarding the unreliable colors $\log
(F_{Kn}/F_{Hn})$ and $\log (F_{Hn}/F_{Jn})$. Tori of relatively small
sizes ($r_m/r_0\sim100$) can give realistic MIR colors and large enough
values of $\log(F_N/F_{Ln})$ for high values of $\tau_e$. Tori of
intermediate sizes ($r_m/r_0$) do not fit the color $\log(F_N/F_{Ln})$
very well. Very extended tori can again well fit all colors. The inclusion
of silicate depression does not change substantially the aforementioned
considerations, although it tends to produce somewhat lower values of
$\log(F_N/F_{Ln})$ for very extended tori (for which, however, the
assumption of uniform density distribution is likely to be less
reasonable). But we have verified that the compatibility between the
theoretical and observed values of $\log(F_N/F_{Ln})$ for very extended
tori critically depends on the adopted degree of silicate depression.  A
lower degree of depression (i.e., a depression within $r\leq15 r_0$) gives
great enough values of that color. 

To sum up, for S2s the two fit parameters are poorly constrained;  the
data clearly rule out only very compact tori ($r_m/r_0\sim$10), which
would require unreasonably large values of $\tau_e$. Within the
unified schemes of AGNs, we would prefer a solution involving very
extended tori also for S2 galaxies. Although the theoretical colors do not
unambiguously converge towards this solution for S2s, it is encouraging
that this possibility is not seriously contradicted. 

The comparison of the observed degree of anisotropy of the IR emission
with the degree predicted by GD models provides a better constrain on
$\tau_e$ than that derived above.  In Fig. 7 we show these predictions
for some photometric bands by plotting the ratio between the fluxes
received by a polar and by an equatorial observer as a function of $\tau_e$,
for different values of $r_m/r_0$. The horizontal lines denote the median
values (i.e. 1.5, 2.7, 6 for the 12 \m-, L-, K-bands, respectively) that
we derived in \S 3.2 by comparing the IR luminosities of S1 and S2 nuclei.
Models with a standard silicate abundance (left column) and with a
standard depression of silicate grains (right column) in the innermost
regions (for $r\leq50 r_0$) yield similar results for moderate values of
$\tau_e$. These plots favor values of $\tau_e$ ranging from $\sim$20 to
$\sim$40 (i.e. 10 mag$\lesssim A_V \lesssim$30 mag) for extended tori
(with $r_m/r_0\gtrsim100$).

\section{Summary and Conclusions}

We have undertaken a statistical analysis of the MIR and NIR fluxes of an
extended sample of Seyfert galaxies, the widest Seyfert sample so far
considered in the relevant literature. We have paid particular attention
in selecting homogeneous data (when available) and nuclear data, i.e fluxes
that are, as far as possible, free of stellar light contamination. This has 
enabled
us to derive the NIR and MIR continuum emission properties of Seyfert
nuclei with unprecedented accuracy. We can summarize our main results as
follows. 

The color distributions indicate that the typical SED  
of S2 nuclei is always steeper than that of S1 nuclei in the whole 
MIR and NIR spectral ranges. Hence S2s become typically fainter with 
respect to S1s as we go towards shorter IR wavelengths.

There is no sure evidence of bimodality in the color and luminosity
distributions. This agrees with the absence of significant bimodality in
the distributions of several emission properties of S1 and S2 galaxies,
such as the ultraviolet, soft X-ray, hard X-ray emissions (Mulchaey et al.,
1994) and radio core emission (Giuricin et al., 1996). A bimodality in
some properties of S2s might be expected if there were two types of S2s,
S2s with obscured BLR and without BLR, as was proposed by some authors 
(e. g., Hutchings \& Neff, 1991; Moran et al., 1992). 

The comparison of the IR luminosities of S1s and S2s,
after they are made free of distance selection effects, reveals that S2s
are somewhat brighter than S1s in the IRAS $\lambda\sim$ 25 \m bands and
in the IRAS FIR bands for the overall sample (though not for all 
subsamples). A direct comparison of the IR luminosities, 
normalized to those of the [OIII] emission line,  confirms these
results. This finding, which is unexpected within the framework of the
unified model of Seyfert galaxies, can be explained in terms of
observational selection effects favoring the inclusion of S2s with
ongoing star formation in several current Seyfert samples, which are 
often dominated by optically--selected objects.

As we go to shorter wavelengths, S2s become increasingly fainter than S1s
by an amount which is in fairly good, quantitative agreement with what we
expect from the differences between their SEDs.  Within the framework of
torus models, this behavior is consistent with an increasing anisotropy
of a torus emission towards shorter IR wavelengths. But the degree of
anisotropy we derive is quite low (i.e. lower than what has frequently
been claimed in the literature). 

For a small sample of S2 galaxies having Compton-thin obscuring structures
at hard X-ray energies, we discover significant correlations (previously
unexplored) between the absorbing hydrogen columns derived from their hard
X-ray spectra and some IR colors and luminosities. This indicates that the
IR emitting torus is somehow related to the X-ray absorbing structure.

The typical shapes of the IR SEDs of S1s and S2s and the luminosity
differences between the two types of objects (i.e. the degree of
anisotropy of the IR radiation) are compared with the predictions of a
wide series of torus models. Contrary to earlier claims, many
models, in particular those involving compact or very thick
configurations, are found to be basically inconsistent with the
observational data. 

We have specifically inspected the basic version of the torus models by
Granato \& Danese (1994), which was intended to represent the simplest
kind of tori envisaged by the AGN unification schemes. We have found that
the major IR observational properties can be reasonably satisfied by
flared disk configurations characterized by quite a large extension (a few
hundreds pc) and moderate equatorial optical thickness (corresponding to a
typical extinction of 10 mag$\lesssim A_V \lesssim$30 mag), for both S1
and S2 galaxies. 

For a standard dust over gas ratio, the aforementioned range of torus
optical thickness, derived from fitting the IR properties of Seyfert
nuclei, translates into lower values of $N_H$ ($N_H\sim2\cdot
10^{22}-5\cdot 10^{22} cm^{-2}$) than the median value of
$N_H\sim2.9\cdot 10^{23} cm^{-2}$ of a sample of S2 galaxies, which is
still probably biased towards low values of $N_H$.  This lends further
support to the recent view that the matter responsible for the X-ray
absorption is mostly dust--free and that the bulk of the X-ray absorption
does not occur in the dusty torus, but at and just inside the dust
sublimation radius (Granato et al., 1997). 

Extended tori are consistent with the detection of dusty nuclear disks on
scales of $\sim$100 pc in some {\it Hubble Space Telescope} observations
(e.g., Jaffe et al., 1993) and with the detection of dense
molecular nuclear disks on similar scales in some high--resolution radio
observations (e. g., Tacconi et al., 1994;  Kohno et al., 1996). Extended
tori are also consistent with the presence of red (i.e. with $V-R>1.0$
mag) regions of hundreds of pc which coincide with the optical
nuclei of many S2 objects; they can be interpreted as being due to
reddening by dusty tori (Mulchaey et al., 1996b). It is known that there is a 
selection bias against finding AGNs (especially UV-excess AGNs and
spectroscopically--selected AGNs) in highly inclined galaxies. This has led
some authors to suggest that the obscuring material covers a part of the
NLR (over $\gtrsim$100 pc) and is roughly coplanar with the host galaxy
disk (McLeod \& Rieke, 1995; Maiolino \& Rieke, 1995). 
 
Extended configurations are also in better agreement than compact tori
(with a radial size of $\lesssim$10pc) with the high--resolution NIR and MIR
observations of NGC 1068, which showed an emission extended over
$\sim$1" scale (e.g., McCarthy et al., 1982; Chelli et al., 1987; 
Tresch--Fienberg et al., 1987; Braatz et al., 1993; Cameron et al., 1993).
If this emission were due to torus radiation reflected by electron
scattering and to NLR dust (Pier \& Krolik, 1993), there would probably be
too large reddening in the NLR (Cameron et al., 1993). The analysis of NIR
polarized images of the nuclear region of this galaxy points to a  torus
with a size of $\sim$200 pc (Young et al., 1996b).  An extended ($\sim$2")
bar-like NIR emission, which might hint at an extended dusty torus
viewed approximately edge--on, has been recently detected in Mk 348
(Simpson et al., 1996). 

Tori with moderate optical depths are consistent with the detection of
broad components of NIR lines in several narrow-line AGNs (as discussed by
Granato et al., 1997). In most cases of detections of broad lines in a
seeming narrow-line object, the broad lines are reddened more than the
narrow lines, locating the bulk of the dust responsible for absorbing the
broad-line emission between the BLR and the NLR (e.g., Hill, Goodrich \&
DePoy, 1996). Narrow--line galaxies which do not show broad lines in total
or polarized flux could have obscured regions of scattered radiation or
lines which are thermally very broadened and hence hardly observable (see,
e.g., Young et al., 1996a). Moderate optical depths are also consistent
with the typical extinctions ($A_V\lesssim30$ mag) inferred from the
strengths of the 10 \m silicate absorption features observed in several S2
galaxies (e.g., Aitken \& Roche, 1985;  Roche et al., 1991). 

Alleviating some earlier worries, we can conclude that the general unified
scheme of AGNs is basically compatible with the IR observational data. 
Notably, this scheme can not hold in a strict sense, since the broadness
of the color distributions suggests a large spread in the properties of
obscuring tori, especially for S2s. Other arguments against the strict
version of the unified scheme can be found in Antonucci's (1993) review.

There is no indication in our sample that the  optical depth of tori
decreases with AGN luminosity, as is advocated by the simplest
modification of the strict unified scheme (e.g., Lawrence, 1991). By the
way, the SEDs of some hyperluminous high--redshift IR galaxies have been
recently reproduced in terms of a luminous AGN surrounded by an extended
torus with a larger equatorial optical depth (60 mag $\lesssim A_V
\lesssim$ 300 mag) than those of classical Seyfert galaxies (Granato,
Danese \& Franceschini, 1996). The ISO satellite observations are expected
to substantially improve our knowledge of the IR properties of
high--luminosity AGNs. 
  
Nevertheless, our detailed comparison with the observational data raises
several problems. Also the most successful torus models, i.e.
extended and moderately thick flared disks, fail to reproduce accurately
several observational details. Even with the present quality and quantity
of observations there is a need for more sophisticated versions of torus
models having more free parameters to fit the data (e.g., different
matter distributions and geometries), though not so many as to preclude
physical understanding. Remarkably, especially in the case of extended
configurations, the possible clumpy nature of the actual dust distribution
weakens the validity of current torus models, which are based on a
continuous distribution of matter.

\acknowledgments
The authors are grateful to L. Danese for enlightening discussions on torus
models and to M. Mezzetti for comments on the statistical analysis.

The authors thank E. D. Feigelson for having kindly provided 
the ASURV software package. 

This research has made use of the NASA/IPAC 
Extragalactic Database (NED), which is operated by the Jet Propulsion 
Laboratory, Caltech, under contract with the National Aeronautics and 
Space Administration. 

This work has been partially supported by the 
Italian Ministry of University, Scientific, and Technological Research 
(MURST) and by the Italian Space Agency (ASI).

\end{multicols}

\end{document}